\documentclass[aip,jcp,reprint,twocolumn]{revtex4-1}
\usepackage{graphicx}
\usepackage{amsmath}
\usepackage{color}
\usepackage[colorlinks=true, linkcolor=black, urlcolor=blue, citecolor=blue]{hyperref}

\newcommand{\beq}{\begin{equation}}
\newcommand{\eeq}{\end{equation}}
\newcommand{\bea}{\begin{eqnarray}}
\newcommand{\eea}{\end{eqnarray}}

\begin{document}

\title{Energetic preference and topological constraint effects on the formation
of DNA twisted toroidal bundles}

\author{Nhung T. T. Nguyen}
\affiliation{Institute of Physics, Vietnam Academy of Science and Technology,
10 Dao Tan, Ba Dinh, Hanoi 11108, Vietnam}

\author{Anh T. Ngo}
\affiliation{Chemical Engineering Department, University of Illinois at
Chicago, Chicago, IL, 60608, USA}

\author{Trinh X. Hoang}
\email[Corresponding author, E-mail: ]{txhoang@iop.vast.vn}
\affiliation{Institute of Physics, Vietnam Academy of Science and Technology,
10 Dao Tan, Ba Dinh, Hanoi 11108, Vietnam}

\affiliation{Graduate University of Science and Technology, Vietnam Academy of
Science and Technology, 18 Hoang Quoc Viet, Cau Giay, Hanoi 11307, Vietnam}

\begin{abstract}
DNA toroids are compact torus-shaped bundles formed by one or multiple DNA
molecules being condensed from the solution due to various condensing agents.
It has been shown that the DNA toroidal bundles are twisted. However, the
global conformations of DNA inside these bundles are still not well understood.
In this study, we investigate this issue by solving different models for the
toroidal bundles and performing replica-exchange molecular dynamics (REMD)
simulations for self-attractive stiff polymers of various chain lengths.
We find that a moderate degree of twisting is energetically favorable for
toroidal bundles, yielding optimal configurations of lower energies than in 
other bundles corresponding to spool-like and constant radius of curvature
arrangements. The REMD simulations show that the ground states of the stiff
polymers are twisted toroidal bundles with the average twist degrees close to
those predicted by the theoretical model. 
Constant-temperature simulations show that twisted toroidal bundles
can be formed through successive processes of nucleation, growth, quick
tightening and slow tightening of the toroid, with the two last processes
facilitating the polymer threading through the toroid's hole.
A relatively long chain of 512 beads 
has an increased dynamical difficulty to access the twisted bundle states
due to the polymer's topological constraint.
Interestingly, we also observed significantly twisted toroidal bundles with
a sharp U-shaped region in the polymer conformation. It is suggested that
this U-shaped region makes the formation of twisted bundles easier by
effectively reducing the polymer length. This effect can be equivalent to
having multiple chains in the toroid. 
\end{abstract}

\maketitle

\section{Introduction}

In DNA condensation, extended DNA chains collapse into highly compact
structures, which contain only one or a small number of molecules
\cite{Bloomfield,Hud05}. The condensation can occur spontaneously in vitro upon
adding a small amount of multivalent cations, such as spermidine$^{3+}$, to a
buffered solution of low ionic strength \cite{Gosule}. It can also be observed
by using polymeric osmolytes such as small peptides or PEG as the condensing
agents \cite{RP1}. 
The packing of DNA inside a condensate is highly ordered and is akin to a
nematic liquid crystalline state \cite{Maniatis}. 
The most commonly observed morphologies of DNA condensates are toroid and
rod-like \cite{Vilfan06}. The sizes of these structures depend on the
solution condition and range from few ten to
few hundred nanometers \cite{Conwell}.
Under certain method of preparation, larger condensates with the spheroid
and V-shaped morphologies can be observed \cite{Pinto}.
Surprisingly, the size of DNA condensates does not depend the contour length
of the DNA molecules involved in the condensation \cite{Bloomfield1990}. 

The phenomenon of DNA condensation has been considered theoretically from the
perspective of the collapse of semiflexible polymers into compact structures,
often in the form of toroidal and/or rod-like globules. Different aspects of
this polymer collapse have been studied in order to deduce the detailed
geometry of the condensates and the resulting phase diagram
\cite{Grosberg1986,Vasilevskaya1997,Yoshikawa1998,Ivanov2000,Stukan2003,Hoang2014,Barbi2015},
their dependence on the forms of the elastic potential
\cite{Stukan2006,Hoang2015} and the DNA-DNA interaction potential
\cite{Ishimoto2008},
as well as on the nature of the collapse transition
\cite{Vasilevskaya1995,Yoshikawa1996} and kinetic pathways
\cite{Yoshikawa2002,MacKintosh2004,Muthukumar2005,Reddy2017}.

There has been much focus on the structural organization of DNA in the toroidal
condensates. In a DNA toroid, the DNA winds around the toroid main axis making
a tight circular bundle. The lateral packing of the filaments inside the
toroid is predominantly hexagonal \cite{Schellman84} but the structure also
contains non-hexagonal parts \cite{Hud2001}. Using cryoelectron microscopy,
Leforestier and Livolant have shown that the DNA toroidal bundle inside the
bacteriophage capsid is twisted with a number of twist walls, the portions of
the toroid in which the hexagonal lattice is rotated, separating non-rotated
hexagonal domains \cite{Livolant2009}. Earlier indirect indications of a
twisted state come from experiments \cite{Marx1983,Hud1995} and simulations
\cite{Stevens2001}, which show that the path of DNA inside the toroids do not
follow the spool-like packaging, and also from the strong knotting of DNA in
phage capsids \cite{Roca2002}. Based on experimental observation, Hud et al.
have proposed a constant radius of curvature model \cite{Hud1995} for the
organization of DNA in the toroids, which is akin to a twisted bundle.  A more
elaborate model of a twisted bundle has been analyzed by Kuli\'c et al. showing
that twisting can spontaneously relax bending energy of the toroidal bundle
\cite{Kulic2004}.  However, the formation of such twisted bundles by a single
or multiple DNA chains in a toroid is still not well understood, which is part
of a more general problem of packing of twisted filament bundles
\cite{Grason2015}.

The present study is aimed at better understanding of the twisted
state of toroidal bundles formed by semiflexible polymer. We focus on
whether the twisted bundles are competitive in energy among several kinds of
bundle organization, such as in the constant radius of curvature model,
and whether such a twisted state is the ground state of a stiff
polymer chain with self-attraction. To get insights into these issues, we
employ a combined approach of using both theoretical models and
coarse-grained molecular dynamics simulations, allowing comparison between
the two methods. We consider two slightly different models for the twisted
toroidal bundles, one with an uniform and another with a variable degree of
twisting, to compare with other models corresponding to spool-like and constant
radius of curvature arrangements. The simulations of the stiff polymers are
carried out both with the REMD method \cite{Sugita1999} for finding low
energy conformations, and with constant temperature molecular dynamics for
studying the toroid formation process.

We will show that a moderate twisting significantly increases the stability 
a toroidal bundle, while the topological constraint of a long polymer
may affect its ability to form a twisted bundle. Interestingly, 
the results obtained from the simulations are in good
agreement with the theoretical models.

\section{Models and Methods}

\subsection{Twisted bundle models}

\begin{figure}
\includegraphics[width=3.4in]{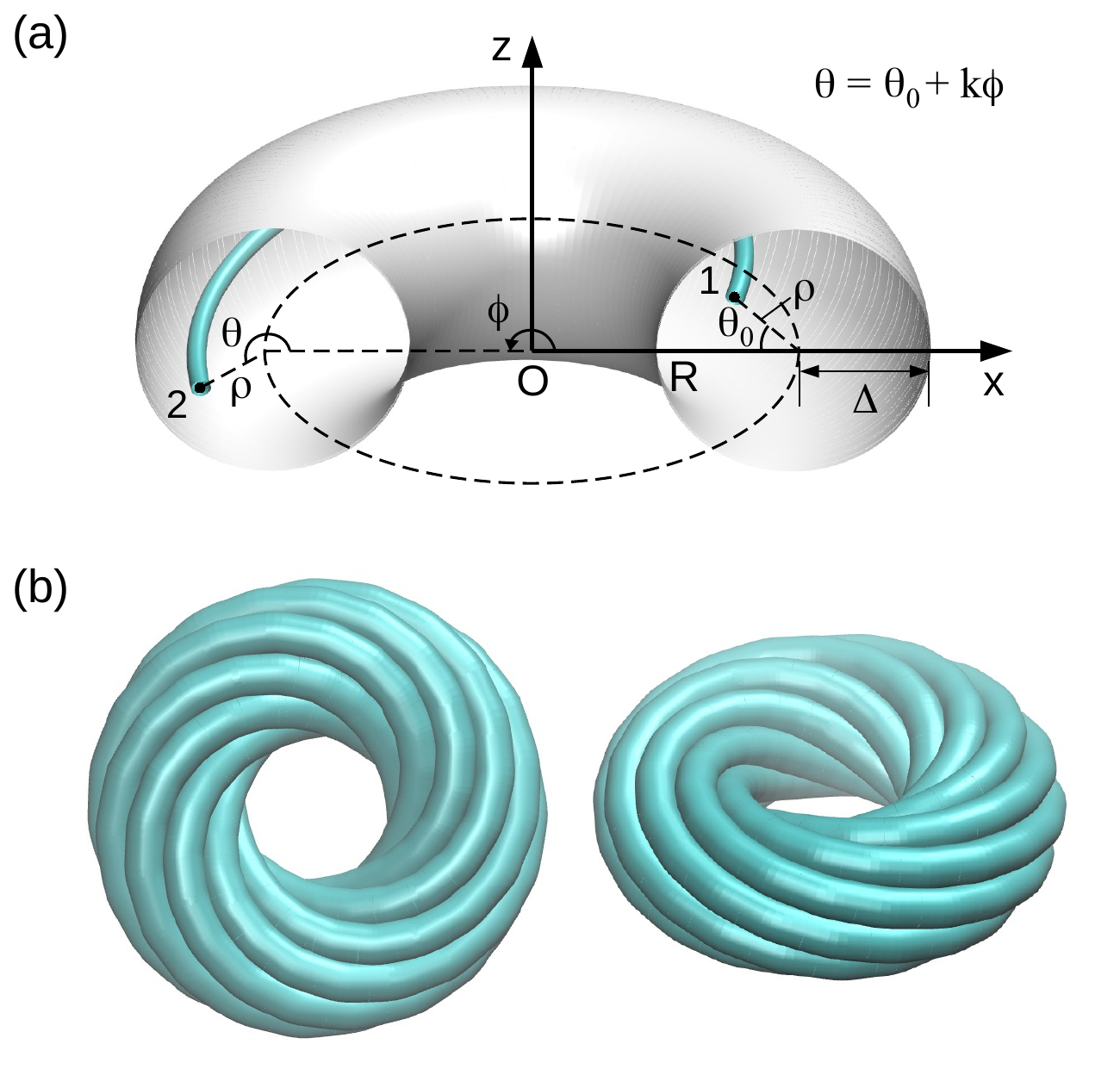}
\caption{(a) Sketch of the coordinates of a toroidal bundle. The toroid's
tubular axis (dashed circle) is set to be centered at the origin $O$ of the
Cartesian coordinates on the $xy$ plane. A DNA filament is shown from point
$1$ to point $2$ inside the torus (only half of which is shown). The filament
trace is determined by the radial distance $\rho$ from the tubular
axis, the rotation angle $\phi$ around the main axis $z$ and the rotation angle
$\theta$ within the tubular cross-section.
$R$ and $\Delta$ are the toroid's mean radius and thickness radius, respectively.
For the twisted toroidal bundle, $\theta = \theta_0 + k\phi$, where $\theta_0$
is a $\theta$'s value at $\phi=0$ and $k$ is the twist number.
(b) Illustration of a twisted toroidal layer formed by a single chain
with $k=0.73$ at two different viewing angles.}
\label{fig:tbmodel} 
\end{figure}

In a twisted toroidal bundle within a toroid of mean radius $R$ and thickness
radius $\Delta$ as shown in Fig.~\ref{fig:tbmodel} (a), the DNA conformation has the
parameterized form \cite{Kulic2004}
\begin{equation}
{\bf r} (\rho,\theta,\phi) = \left(
\begin{array}{c}
\left(R - \rho \cos \theta \right)\cos \phi \\
\left(R - \rho \cos \theta \right) \sin \phi \\
\rho \sin \theta
\end{array}
\right) 
\label{eq:r}
\end{equation}
with $\theta=\theta_0 + k\phi$ and the column on the right contains the
$x$, $y$ and $z$ components of the vector ${\bf r}$. 
In this parameterization, ${\bf r}$ is the
position of the DNA molecular axis, $\rho \in
(0,\Delta)$ is the radial distance from the toroid's tubular axis, $\phi$ and
$\theta$ are the rotation angles
around the main and tubular axes, respectively, and $k$ is a parameter which
sets the degree of twisting. $k$ defines how much the change in $\theta$ is
faster than the change in $\phi$ along the filament, and hereby called the twist
number. Such a bundle is organized into disconnected toroidal layers of
filaments at constant $\rho$ as shown in Fig.~\ref{fig:tbmodel} (b).

Denote ${\bf r}^\prime = d{\bf r}/d\phi$. It can be shown that
\begin{equation}
|{\bf r}^\prime| = \sqrt{(k \rho)^2 + \left(R - \rho \cos \theta \right)^2} \ ,
\end{equation}
which gives the differential of the arc length $s$ of a filament,
$ ds = |{\bf r}^\prime| d\phi $.
The tangent ${\bf t}$ is defined as
\begin{equation}
{\bf t} = \frac{d{\bf r}}{ds} 
=\frac{1}{|{\bf r}'|} \frac{d{\bf r}}{d\phi} 
\ . 
\end{equation}
The normal ${\bf n}$ and the binormal ${\bf b}$ are defined through the
Frenet-Serret equations \cite{Kreyszig}:
\begin{equation}
\frac{d{\bf t}}{ds} = c\, {\bf n} \qquad 
\frac{d{\bf n}}{ds} = - c\, {\bf t} + \tau_s {\bf b} \qquad
\frac{d{\bf b}}{ds} = -\tau_s {\bf n}
\label{eq:FS}
\end{equation}
which also define the curvature $c$ and the torsion $\tau_s$ of the
DNA axis. From the parameterization in Eq.~(\ref{eq:r}), $c$ can be
obtained in a closed form, whereas $\tau_s$ is more conveniently calculated
numerically. Due to the toroid symmetry, both $c$ and $\tau_s$ are functions
of $\rho$ and $\theta$, only.

Given that $d$ is the lateral distance between neighboring DNA filaments
and $\eta = \pi/(2\sqrt{3})$ is the volume fraction of a hexagonal packing,
the total length of DNA in the toroid is given by
\begin{equation}
L=\frac{\eta V_\mathrm{tor}}{\pi (d/2)^2} 
= \frac{8\pi \eta \Delta^2 R}{d^2} \ .
\end{equation}
One can write the bending energy of the toroidal bundle in the volume integral
form
\begin{equation}
U_\mathrm{tor} = 
\frac{\eta}{\pi(d/2)^2} \int_0^\Delta d\rho \int_0^{2\pi} \rho\, d\theta \,
\int_0^{2\pi} (R-\rho \cos\theta) 
\,\frac{A}{2} c^2 \, d\phi \ ,
\label{eq:Utor}
\end{equation}
where $A$ is the DNA bending stiffness \cite{Marko1995}. Note
that this form of bending energy implies an uniform density of DNA inside the
toroid, or equivalently, a constant spacing $d$ within the toroid. The latter
is an approximation since a recent experiment has shown that this
spacing decreases with increasing distance from the toroid center
\cite{Barberi2021}.

We adopt an energy function for
the toroid, which includes only bending and surface energy terms:
\begin{equation}
E_\mathrm{tor} = U_\mathrm{tor} + \sigma S_\mathrm{tor} \ ,
\end{equation}
where $\sigma$ is the surface tension and $S_\mathrm{tor} = 4 \pi^2 \Delta R$ is
the toroid's surface area. 
The present model neglects the twist degree of freedom of the DNA double
helix and also assumes that the chain is inextensible. These are reasonable
approximations given that the molecule is not under torque nor tension. Indeed,
it can be shown that the twist energy in toroids due to the
twist-bend coupling \cite{Marko1994} is negligible (see Discussion).
However, it should be noted that our model also neglects the helical
structure of DNA, which may give rise to the twisting of DNA bundles
\cite{Livolant2009}.

Let us call $\alpha = \Delta/R$ the thickness radius to mean radius ratio
or shortly the thickness ratio.
The equilibrium configuration of the toroid is obtained by minimizing the
energy with respect to $\alpha$ and $k$.
Hence, $E_\mathrm{tor}$ plays the role of a free energy at 
a constant temperature.
In numerical calculations, we set $d=2.8$~nm, and the DNA bending stiffness 
$A=50~\mathrm{nm}\cdot k_B T$ with $k_B$ the Boltzmann constant and $T$ the
absolute temperature. Energy is given in units of $k_B T$. 

We consider two models of twisted toroidal bundles. In the first model, namely
the twisted bundle (TB) model, the twist number $k$ is independent of $\rho$ and
thus is uniform in the whole toroid. This model is identical to the
conventional model considered elsewhere \cite{Kulic2004}. The second model
allows $k$ to be varied with $\rho$ and thus is hereby named the twisted bundle
model with a $\rho$-dependent twist number (TB-$\rho$ model). In both models,
$k$ is optimized by minimizing the toroid energy.

\subsection{Spool-like model}

The spool-like (Sp) model \cite{Hud1995} is a special case of the TB model with
$k=0$.  In the Sp model, the curvature is given by
\beq
c = \frac{1}{R-\rho\cos\theta} \ .
\eeq
and the bending energy in Eq. (\ref{eq:Utor}) can be exactly calculated giving
\begin{equation}
U_\mathrm{tor}^\mathrm{Sp} = 
\frac{8 \pi \eta A}{ d^2}\, (R - \sqrt{R^2 - \Delta^2}) \ .
\end{equation}

\subsection{Constant radius of curvature model}

The constant radius of curvature (CC) model \cite{Hud1995} sets
an uniform curvature of DNA within the toroid with the radius of
curvature equal to the toroidal mean radius $R$. Such a model is formed by
loops being deposited spirally around the toroid circular axis (see Ref.
\cite{Hud1995}). 
The bending energy in this model is equal to
\begin{equation}
U_\mathrm{tor}^\mathrm{CC} = \frac{A}{2} \frac{L}{R^2} \ . 
\end{equation}

\subsection{Stiff polymer model with self-attraction}

We consider also a stiff polymer model for studying the toroid formation 
by molecular dynamics simulations. The model considers a chain of $N$
beads with the potential energy given by
\begin{eqnarray}
E_p &=& \sum_{i=1}^{N-1} K ( r_{i,i+1} - b)^2 
+ \sum_{i=1}^{N-2} \kappa_b (1 - {\bf u}_i \cdot {\bf u}_{i+1})
\nonumber \\
& &
+ \sum_{i<j+2} \epsilon \left[e^{-2(r_{ij}-d_0)/\lambda} 
- 2 e^{-(r_{ij}-d_0)/\lambda} \right] , 
\end{eqnarray}
where the first term contains harmonic potentials for the chain
connectivity with the equilibrium bond length $b$ and spring
constant $K$, the second term is a worm-like chain or Kratky-Porod's type of
bending energy \cite{Marko1995,Rosa2003} with $\kappa_b$ the stiffness per
bead, and the last term corresponds to non-local interactions given by
the Morse potential \cite{Hoang2014} with the potential depth $\epsilon$, the
equilibrium length $d_0$ and the decay length $\lambda$;  
$r_{i,i+1}=|{\bf r}_{i+1} - {\bf r}_{i}|$ is 
the distance between beads $i$ and $i+1$; ${\bf u}_i$ is
a normalized vector given by
\begin{equation}
{\bf u}_i = \frac{{\bf r}_{i+1} - {\bf r}_{i}}{r_{i,i+1}};
\end{equation}
$r_{ij}$ is the distance between bead $i$ and bead $j$. 
All the beads are assumed to have the same mass $m$.
For the simulations, we consider $b$, $m$ and $\epsilon$ as the length,
the mass and the energy units, respectively. 
The parameters chosen for the model are $K=100\,\epsilon/b^2$,
$\kappa_b=22\epsilon$, $d_0=1.4 b$, and $\lambda=0.24 b$. Given that the DNA
thickness is about 2 nm, which
is equivalent to $b$, the chosen values $d_0$ and $\lambda$ are close
to those of the intermolecular potential in multivalent cation condensed DNA
measured by osmotic stress \cite{Todd}.

To analyze the twist degree of a toroidal structure of the polymer obtained by
the simulations, we first fit the polymer conformation to a perfect torus. 
The torus center is the center of mass of the polymer.
The main axis of the torus is determined by diagonalizing the inertia tensor
of the polymer conformation. The mean radius $R$ is obtained by minimizing
the root mean square distance of all the beads of the polymer from the 
tubular axis. A local tangent vector at the bead $i$ of the polymer is defined
as
\begin{equation}
{\bf t}_i  = \frac{{\bf r}_{i+1} - {\bf r}_{i-1}}
{|{\bf r}_{i+1} - {\bf r}_{i-1}|}\ .
\end{equation}
The local twist number of the bead $i$ at the distance $\rho$ from
the tubular axis and with the toroidal rotation angle $\theta$ is calculated as
\begin{equation}
k_i = \frac{(R - \rho \cos \theta)\, t_{\phi} }{\rho \, t_{\theta}} ,
\end{equation}
where $t_{\phi} = {\bf t}_i \cdot {\bf e}_\phi$
and $t_{\theta} = {\bf t}_i \cdot {\bf e}_\theta$
are the $\phi$- and $\theta$-components of the tangent vector, respectively,
with ${\bf e}_\phi$ and ${\bf e}_\theta$ the unit vectors along
the corresponding rotational directions.

\subsection{Replica-exchange molecular dynamics (REMD)}

The replica-exchange molecular dynamics (REMD) method
\cite{Sugita1999} is implemented to find  the ground state of the stiff polymers. In this
method, one simulates multiple copies (or replicas) of a system at various
constant temperatures and
regularly attempts swap moves that exchange the replica conformations at
neighboring temperatures with associated velocity rescaling.
The exchange probabilities are determined such that the detailed
balance condition is satisfied at each temperature \cite{Swendsen1986}. 
This parallel tempering technique has been widely used in molecular simulations
and is very efficient for obtaining equilibrium characteristics as well as
the ground state of a system. In our REMD simulations, the constant temperature
runs for the replicas were carried out by using a molecular dynamics method
based on the Langevin equation \cite{Thuy2016}.
For a given polymer, 16--30 replicas were simulated at a range of temperatures
that spans from a high temperature corresponding to the swollen phase of the
polymer to a low temperature near zero. Swap moves were attempted every
$10\,\tau$, where $\tau = (m b^2/\epsilon)^{1/2}$ is the simulation's time
unit. The lengths of the simulations are of the order of $10^6\,\tau$ for each
replica. For each polymer length, up to 24 independent REMD simulations were
carried out. The ground state is considered to be the lowest energy
conformation if the same conformation was obtained in several runs. 

\section{Results}

\begin{figure}
\includegraphics[width=\columnwidth]{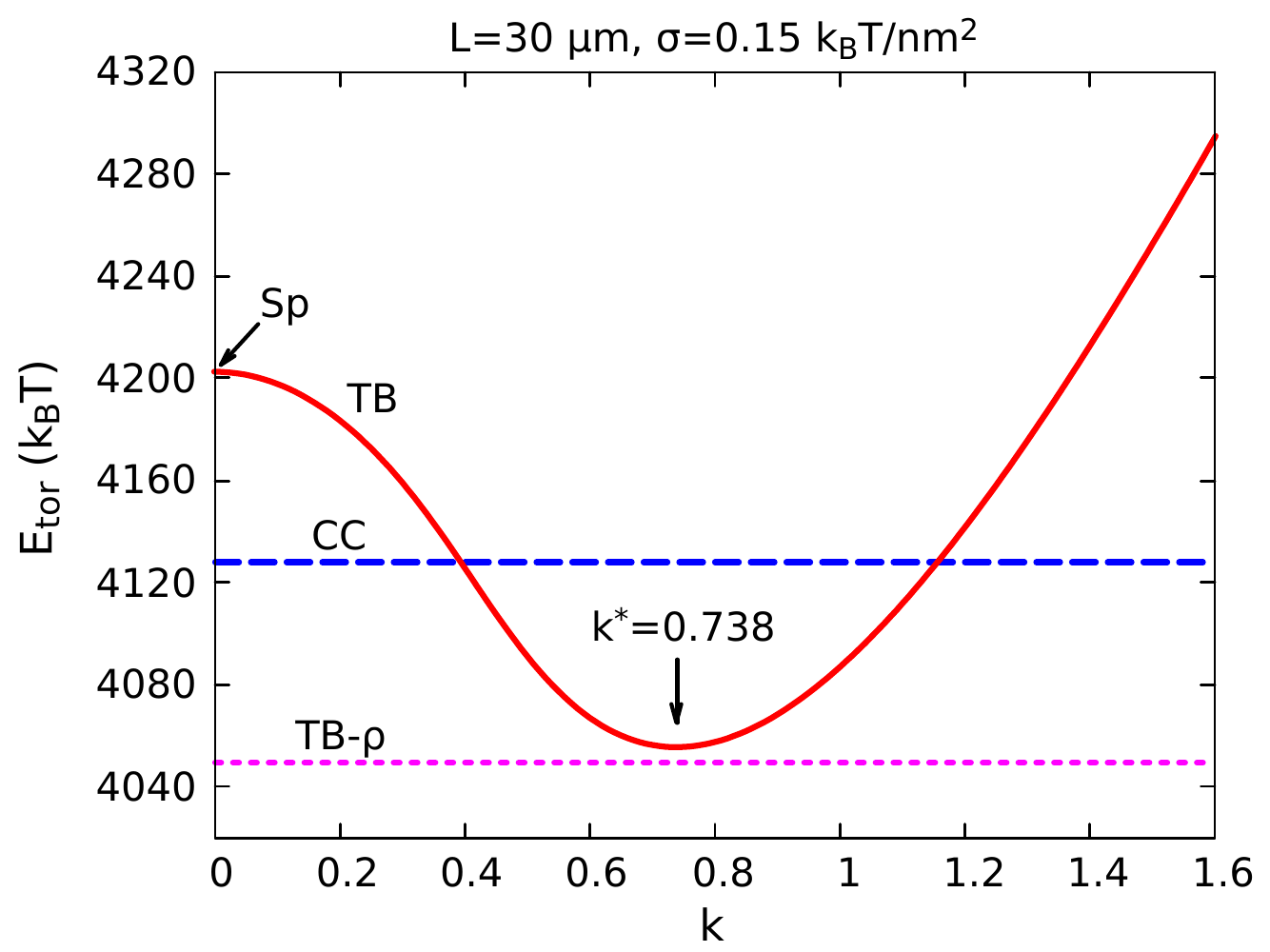}
\caption{Dependence of the toroid energy on the twist number $k$ in the
twisted bundle model (TB) (solid) for the DNA length $L=30~\mu$m and
the surface tension $\sigma=0.15~k_BT/\mathrm{nm}^2$.
For each value of $k$, the energy is minimized with respect to
the thickness ratio $\alpha$. The energy at $k=0$ corresponds to the 
spool-like model (Sp). Horizontal lines indicate the toroid energy
in the constant radius of curvature model (CC)
(dashed) and in the twisted bundle model
with a $\rho$-dependent twist number (TB-$\rho$) (dotted).
}
\label{fig:energy}
\end{figure}

We first studied the characteristics of the 4 models of toroidal bundles,
TB, TB-$\rho$, Sp and CC, described in the Methods section. For a given DNA
length $L$ and a surface tension $\sigma$, the toroid energy is minimized with
respect to the 
thickness ratio $\alpha = \Delta/R$ with $\alpha \in (0,1)$ in all the models.
In the TB and TB-$\rho$ models, the energy is also minimized with respect to
the twist number $k$ and
$k(\rho)$, respectively. Typical toroids have $L$ from 15~$\mu$m to 30~$\mu$m,
whereas a giant toroid can reach $L\sim10^3$~$\mu$m.
The value of $\sigma$ can be estimated from the intermolecular
potential $\Phi(d)$ for DNA measured by osmotic stress experiment \cite{Todd} as
$\sigma=-\frac{\Phi(d)}{d}$ \cite{Hoang2014}.
Consider an example of  $L=30~\mu$m and $\sigma=0.15~k_BT/\mathrm{nm}^2$. 
The TB model has the energy minimum at the optimal twist number 
$k^*\approx 0.738$ as shown in Fig.~\ref{fig:energy}. This energy minimum is
substantially lower than the lowest energy of the toroid in the Sp and the CC
models with the energy difference from $\sim$70 to $\sim$150 $k_BT$. The TB-$\rho$
model yields an energy only $\sim$6~$k_BT$ lower than the TB model and is the one
that gives the lowest energy among all the models. The above picture of the
energy competition is qualitatively the same for all $L$ and $\sigma$,
indicating that twisting can substantially stabilize the toroidal bundle. 

\begin{figure}
\includegraphics[width=\columnwidth]{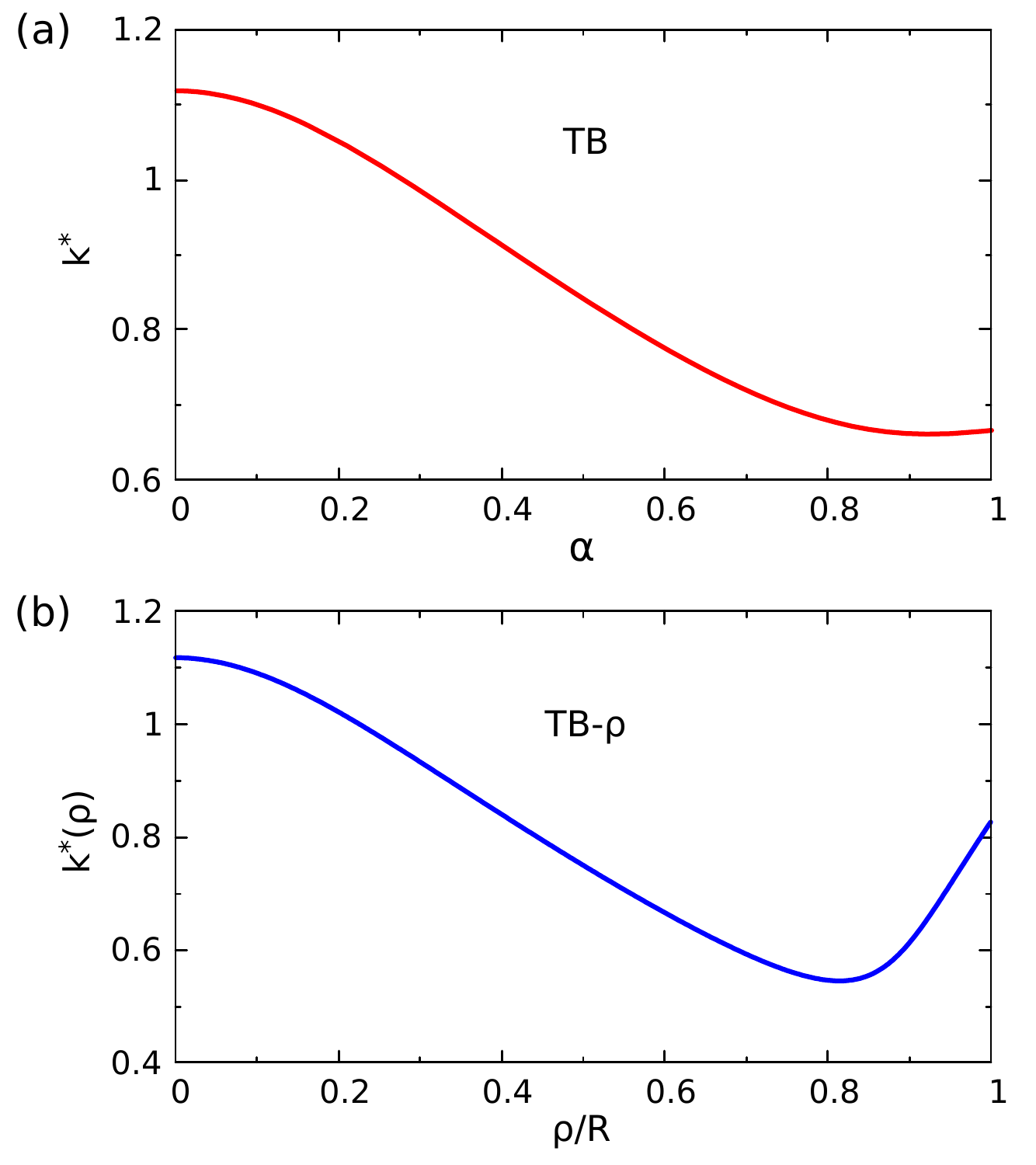}
\caption{(a) Dependence of the optimal value of the twist number, $k^*$, on the
toroid's thickness ratio $\alpha$ in the TB model. (b)
Dependence of the optimal value of the $\rho$-dependent twist number,
$k^*(\rho)$, on the ratio $\rho/R$ in the TB-$\rho$ model. }
\label{fig:kstar}
\end{figure}

Due to the system's geometry, the optimal twist number $k^*$ in the TB model is
a function of the thickness ratio
$\alpha$ alone. As shown in Fig.~\ref{fig:kstar} (a),  $k^*$ is a non-monotonic
but primarily decreasing function of $\alpha$. The values of $k^*$ are
bound between the maximum of about 1.12 obtained at the limit of $\alpha=0$
and the minimum of about 0.66 at $\alpha \approx 0.89$. Similarly, the
optimal twist number $k^*(\rho)$ in the TB-$\rho$ model is a function of
$\rho/R$ alone. Figure \ref{fig:kstar} (b) shows that the dependence of
$k^*(\rho)$ on $\rho/R$ is non-monotonic with a maximum at $\rho/R=0$ and a
minimum at $\rho/R\approx 0.81$. The range of $k^*(\rho)$, from 0.55 to 1.12,
is slightly different from that of $k^*$. The ranges of $k^*$ and $k^*(\rho)$
indicate that twisting is moderate in the optimal toroidal bundles.

\begin{figure}
\includegraphics[width=\columnwidth]{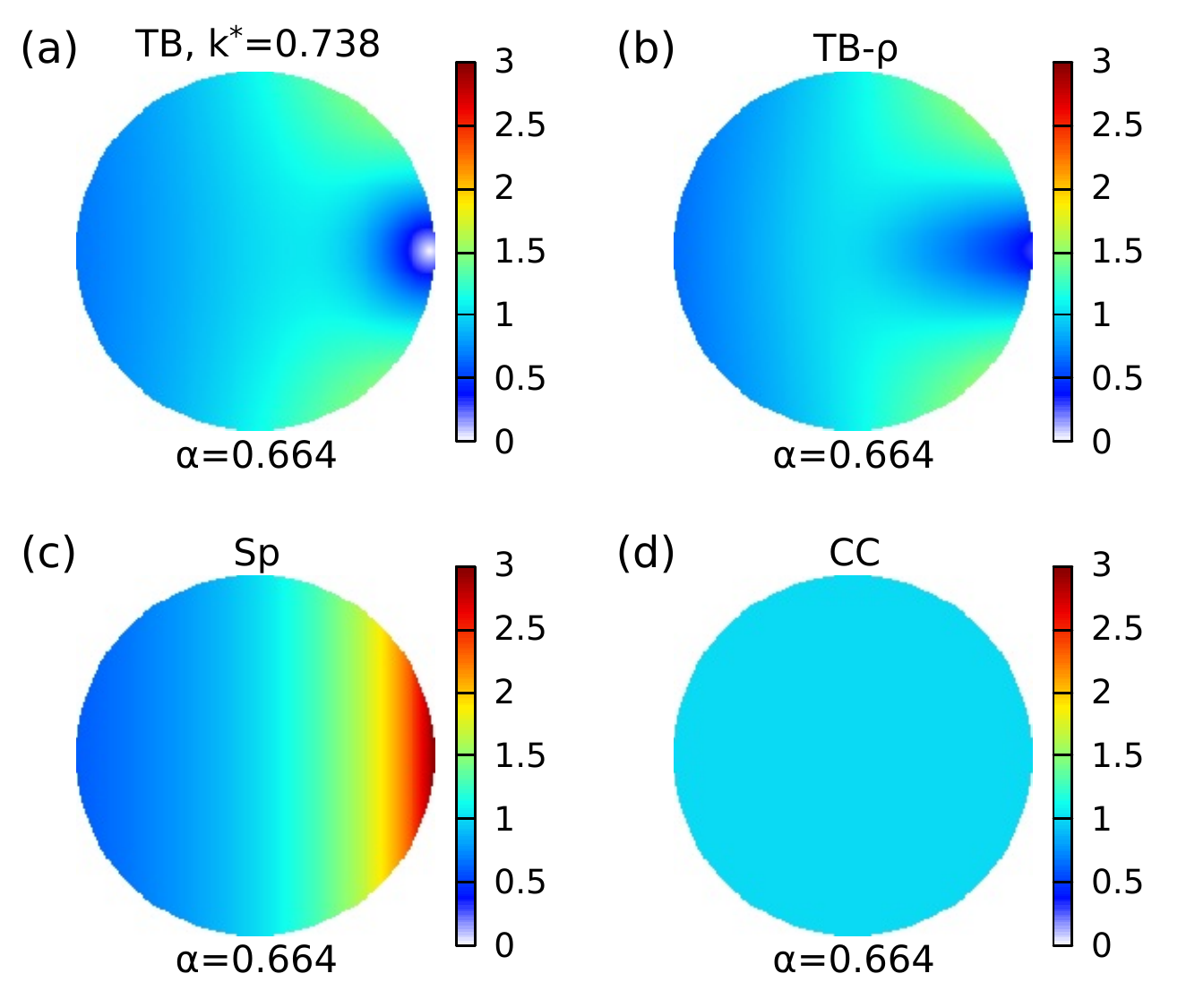}
\caption{Curvatures of DNA in modelled toroids with thickness ratio
$\alpha=0.664$. The curvatures, with values indicated by
the color bar in units of $R^{-1}$, are shown as
color maps for a tubular cross section of the
toroidal bundles in the TB (a), TB-$\rho$ (b), Sp (c), and CC (d) models. 
The toroid in the TB model is shown with the optimal twist
number $k^*=0.738$. The right edge of the cross section corresponds to 
the inner edge of the toroid.
} \label{fig:cmap}
\end{figure}

Figure \ref{fig:cmap} shows the curvature maps in a tubular cross section
of the toroids in the four models considered. These maps, obtained for
$\alpha=0.66$ and suitably at optimal values of $k$ and $k(\rho)$, show that
the TB and TB-$\rho$ models have a broaden spot of low curvatures near the
inner edge of
the toroid (the right edges of the cross sections in Fig. \ref{fig:cmap}).
At this region of the cross section, the curvatures are the highest in the Sp
model. This difference clearly shows the effect of twisting on lowering the
curvatures of the bundles. 
On increasing $\alpha$, we find that the low curvature spot in the TB
and TB-$\rho$ models is slightly shifted away from the inner edge of the toroid
(see Fig. S1 for $\alpha=0.9$) but the qualitative picture is unchanged.
No big difference is seen in the curvature maps of the TB and TB-$\rho$ models,
indicating that these two models work similarly. 

\begin{figure}
\includegraphics[width=\columnwidth]{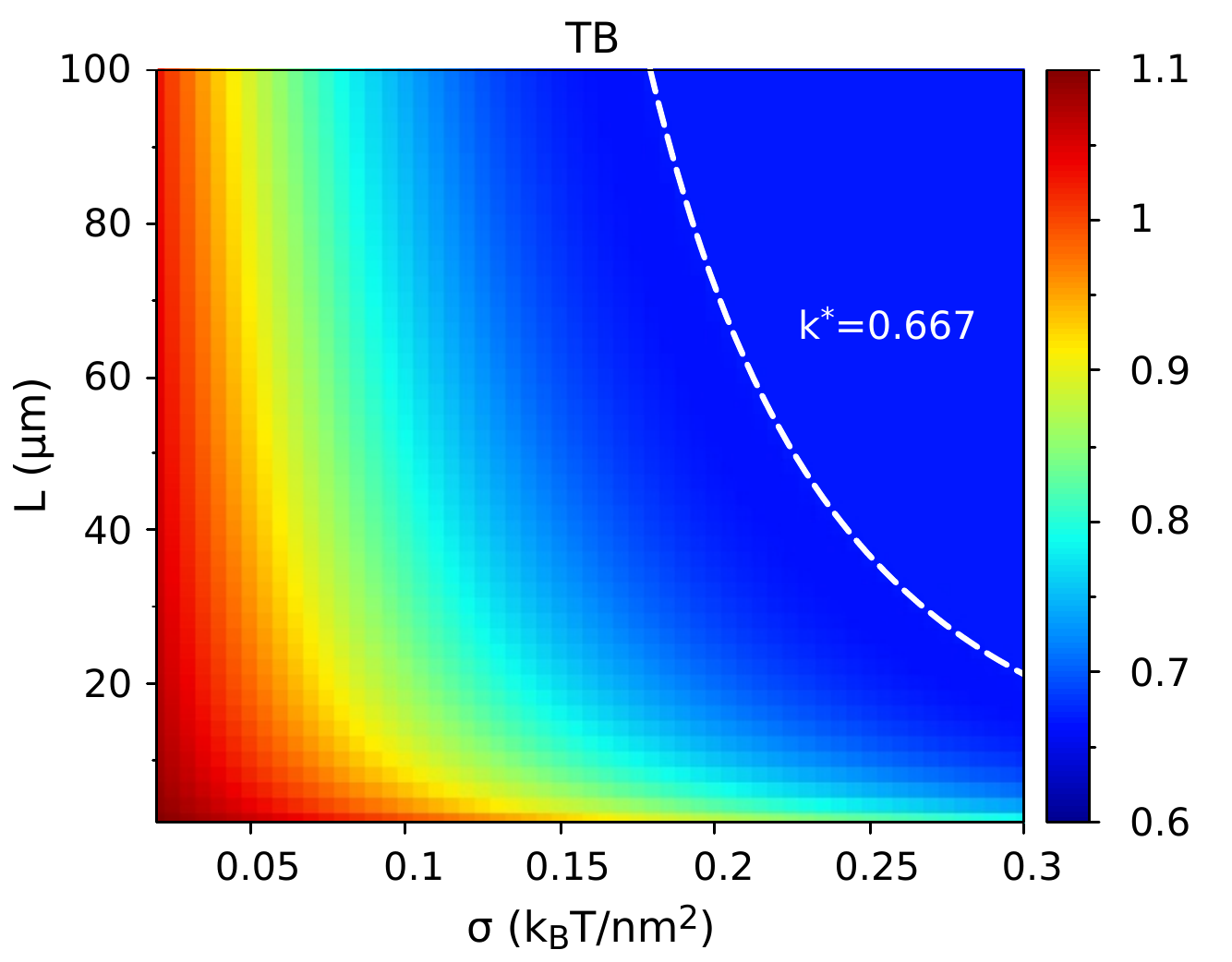}
\caption{Dependence of the optimal twist number, $k^*$, on the DNA length,
$L$, and the surface tension, $\sigma$, of toroidal condensates in the
twisted bundle model. The values of $k^*$ are shown by a heat map with
colors indicated by the associated color bar. Dashed line indicates the
boundary of a map region in which $k^*$ is constant ($k^*=0.667$). In this
region the toroid has no hole ($\alpha=1$). The boundary has the
shape of $L\sim \sigma^{-3}$. }
\label{fig:kmap}
\end{figure}

Figure \ref{fig:kmap} shows the dependence of the optimal twist number, $k^*$,
on the DNA length, $L$, and the surface tension, $\sigma$, as a heat map for the
toroidal bundles in the TB model. It can be seen that that $k^*$ gradually
decreases from the top values to some minimum value on increasing $L$ or
$\sigma$ with the gradient in $\sigma$ much higher than in $L$. There is a
region in the top right of the $\sigma$-$L$ plane where $k^*$ is constant and
equal to 0.667.  This region corresponds to the toroids with $\alpha=1$ or the
toroids with no holes (see also Fig. S2). The boundary of this region is found
to have a shape of $L \sim \sigma^{-3}$ in the $\sigma$-$L$ plane, indicating
an invariance of $\alpha$ and hence $k^*$ by rescaling $\sigma$ with $L^{1/3}$.

\begin{figure}
\includegraphics[width=\columnwidth]{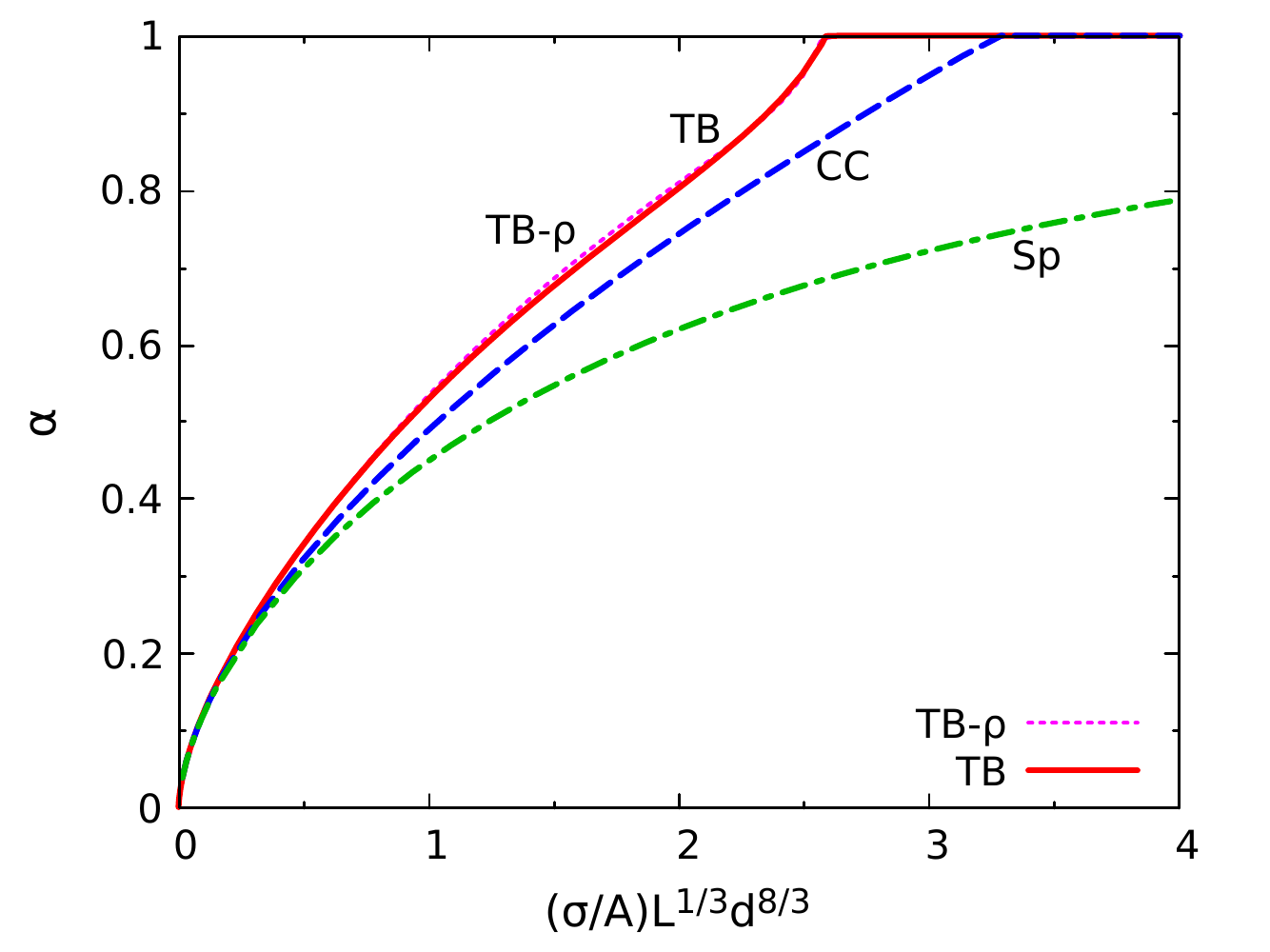}
\caption{Dependence of the thickness ratio $\alpha$ on
the scaled and dimensionless quantity $(\sigma/A)L^{1/3}d^{8/3}$ for
toroidal bundles in the TB (solid), TB-$\rho$ (dotted), CC (dashed), and Sp
(dotted), as indicated. In all models, the toroid energy is minimized with
respect to geometrical parameters.
}
\label{fig:ratio}
\end{figure}

\begin{figure*}
\includegraphics[width=18cm]{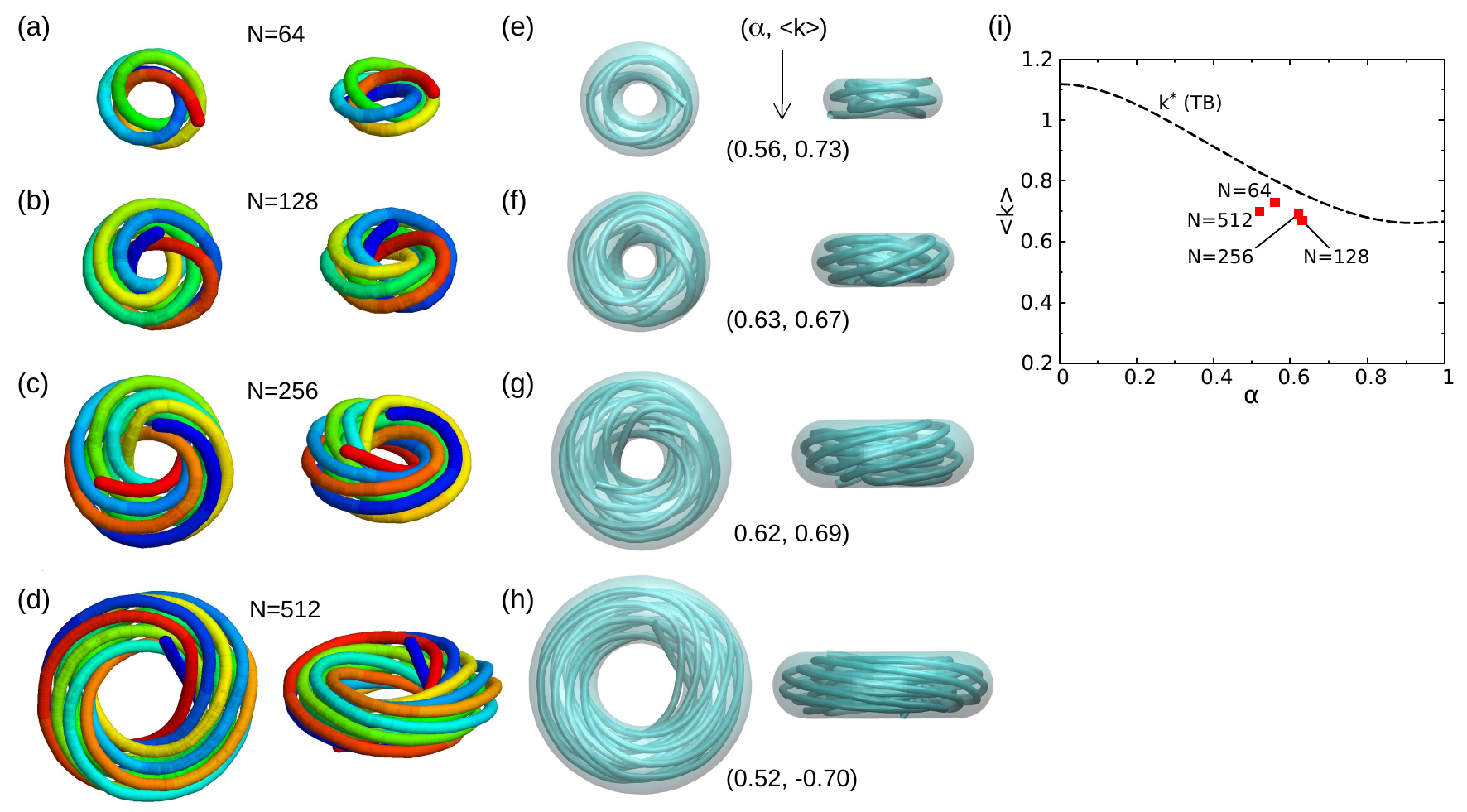}
\caption{(a--d) Toroid conformations obtained by REMD simulations in the
self-attractive stiff-polymer model with $\kappa_b=22\epsilon$. The conformations shown
are the lowest energy conformations for the chain lengths $N=64$ (a) $N=128$
(b), $N=256$ (c) and $N=512$ (d). Each conformation is shown at two different
viewing angles.  (e--h) Fits of the bundles from left into a perfect torus
shown at two different angles.  The analyses of the bundles (see Methods and
Fig. S3) give the estimated values of $\alpha$ and the average twist number,
$\langle k \rangle$, for each conformation (numbers in parentheses).  
(i) Dependence of the average twist number, $\langle k \rangle$,
on the thickness ratio $\alpha$ for the toroidal bundles shown in (a--d)
(squares).  The data points are labeled with the corresponding chain lengths.
For comparison, the dependence of the optimal twist number, $k^*$, on $\alpha$
from the TB model is also shown (dashed).
} \label{fig:md1}
\end{figure*}

\begin{figure}
\includegraphics[width=\columnwidth]{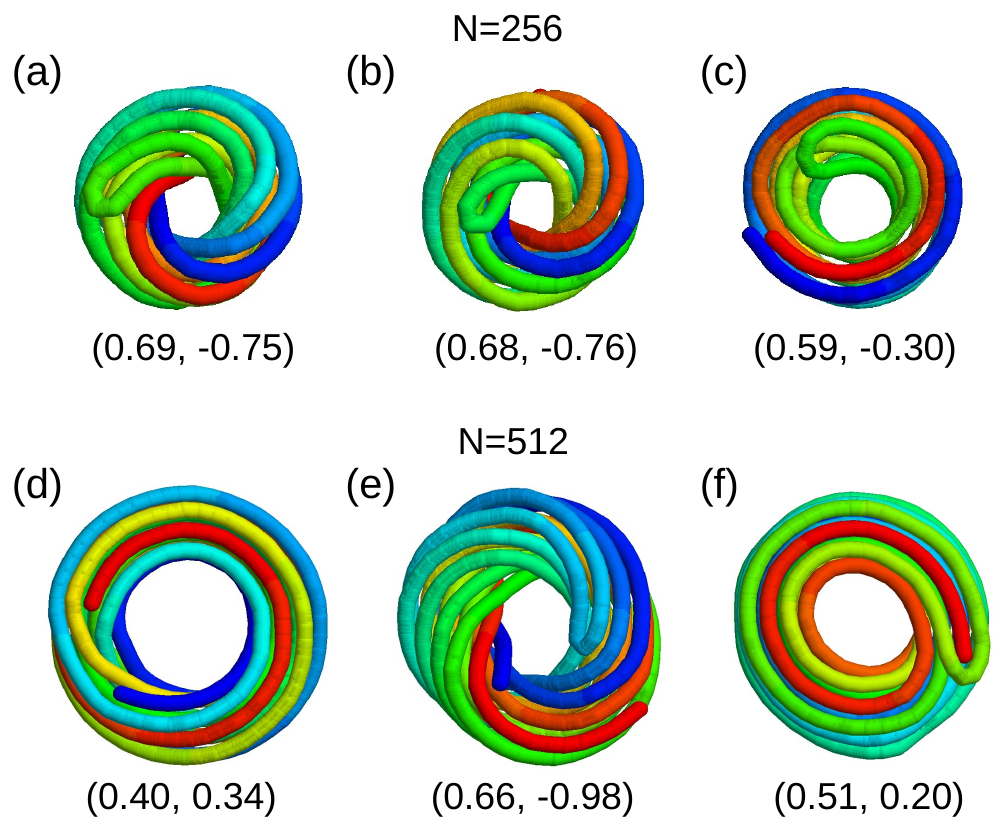}
\caption{Some low-energy toroidal conformations obtained in REMD simulations
for stiff polymers of lengths $N=256$ (a, b, c) and $N=512$ (d, e, f) with
$\kappa_b=22\epsilon$. The energies of these conformations are higher than but within
2\% from the energy of the conformations shown in Fig. \ref{fig:md1}(c and d).
The two numbers in parentheses under each conformation correspond to 
the thickness ratio $\alpha$ and the average twist number $\langle k \rangle$,
respectively, obtained from the analysis of the bundle. The local twist numbers
of the residues in the U-shaped regions are discarded in calculating
$\langle k \rangle$.  }
\label{fig:md2}
\end{figure}

In fact, in all the models considered, the properties of toroidal bundles are
invariant with the scaled quantity $(\sigma/A)L^{1/3}$.  Figure
\ref{fig:ratio} shows that the thickness ratio $\alpha$ increases with
$(\sigma/A) L^{1/3}$ in all the models considered, but
the toroids in the TB and TB-$\rho$ models have higher thickness ratio than in
the CC and Sp models for the same $(\sigma/A) L^{1/3}$. The value of
$\alpha$ therefore reaches 1 faster in the TB and TB-$\rho$ models on
increasing $L$ or $\sigma$. These two twisted bundle models also yield
very similar values of $\alpha$, as shown in Fig.~\ref{fig:ratio}.

Finally, we carried out REMD simulations of stiff polymers with self-attraction
to study the toroid formation.  We fixed the
stiffness per bead of these polymers to be $\kappa_b=22\epsilon$ and studied various
chain lengths of $N=64$, 128, 256 and 512 beads. For the stiffness and length
considered, the simulations show that the polymers form toroids at low
temperatures and rods at intermediate temperatures. 
Figure~\ref{fig:md1} (a--d) shows the lowest energy conformations obtained by the
simulations for these systems, all of which have the toroidal shape.
For $N=64$, 128 and 256, the conformations shown in Fig.~\ref{fig:md1} are very
likely the ground state of the corresponding system as we have obtained very
similar conformations of similar energies in several independent simulations. 
For $N=512$, this is less likely due to the large system size and 
the decreased dynamical accessibility of the lowest energy states (the lowest
energy conformation in Fig.~\ref{fig:md1} (d) was obtained only once).
Interestingly, all these ground state and lowest energy structures appear
as twisted toroidal bundles with the polymer winding repeatedly from the
outer perimeter to the inner hole of the toroid. They look remarkably similar
to the ideal twisted bundle shown in Fig.~\ref{fig:tbmodel} (b).

In order to quantitatively estimate of the twist degrees of the lowest energy
toroidal bundles, we fitted the polymer conformation to a perfect torus
for each system, and then calculated the local twist numbers from the local
tangent vectors of the polymer (see Methods and Fig.~S3). The images of the
fits are shown in Fig.~\ref{fig:md1} (e--h). Our analysis gives the average
twist number $\langle k \rangle \approx 0.73$, 0.67, 0.69 and $-0.7$, for
the $N=64$, 128, 256 and 512 systems, respectively.  The minus sign for the
average twist number of the $N=512$ system stands for an inverse twisting
direction of this bundle compared to the other bundles.
For all the bundles shown in Fig.~\ref{fig:md1}, $\langle k \rangle$
falls within the range of $k^*$ predicted by the TB model. 
By plotting
$\langle k \rangle$ vs. $\alpha$ for the lowest energy toroids obtained by
simulations, as shown in Fig.~\ref{fig:md1} (i), we find that $\langle k
\rangle < k^*$ for all systems considered but the data points
are quite close to the curve of $k^*$ given by the TB model. 
%It is also interesting to notice that the structure of the toroidal
%bundle for the $N=512$ system can be divided into a non-twisted domain and a
%twisted domain, as shown in Fig.~\ref{fig:md1} (h). The appearance of
%the twisted domain is quite similar to the twist walls observed in experiment
%\cite{Livolant2009}.

Note that we have also obtained many competing toroidal conformations from
independent REMD simulations. Some of them are shown in Fig. \ref{fig:md2} for
$N=256$ and $N=512$ systems. The energies of these conformations are only
slightly (less than 2\%) higher than the energies of the lowest energy
conformations shown in Figs. \ref{fig:md1} (c and d). 
Note that though having low energies, some of these toroidal bundles 
are only weakly twisted [Fig.~\ref{fig:md1} (c, d and f)].
Interestingly, some other toroidal bundles are strongly twisted.
For example, the conformations shown in Fig.~\ref{fig:md2} (a, b, and e) 
are even more twisted than the lowest energy conformations in
Fig.~\ref{fig:md1} (c and d). Note that the strongly
twisted bundles in Fig.~\ref{fig:md2} (conformations a, b, and e) have a
sharp U-shaped region in the polymer conformation.
It seems that this U-shaped region makes the polymer effectively shorter and
hence allows easier access to the twisted bundle conformation with some cost in
energy.

The REMD simulations carried out above are efficient for the energy
minimization but they contain unrealistic conformational changes due to the
replica exchange moves.  In order to study the toroid formation process with
true dynamics, we carried out multiple constant-temperature MD simulations of
the stiff polymer of $N=256$ beads at temperatures $T$ below the collapse
transition temperature, which is roughly $\sim$$2.7\,\epsilon/k_B$.  At
$T=0.5\,\epsilon/k_B$ and $T=1\,\epsilon/k_B$, nearly 40\% of the
trajectories succeeded in forming a toroid within the simulation time limit of
$10^6\tau$ for each temperature.  Other trajectories ended up in hairpin-like
and rod-like conformations (see Fig. S4), which are known to be metastable
states of stiff polymers \cite{MacKintosh2004}.  At $T=2\,\epsilon/k_B$,
almost all the trajectories ended up in a toroid within the above time limit.
From these simulations, we find that toroids can be formed from extended
conformations either through a direct pathway (without intermediates) or
through an indirect pathway (with intermediates). 

\begin{figure*}
\includegraphics[width=13cm]{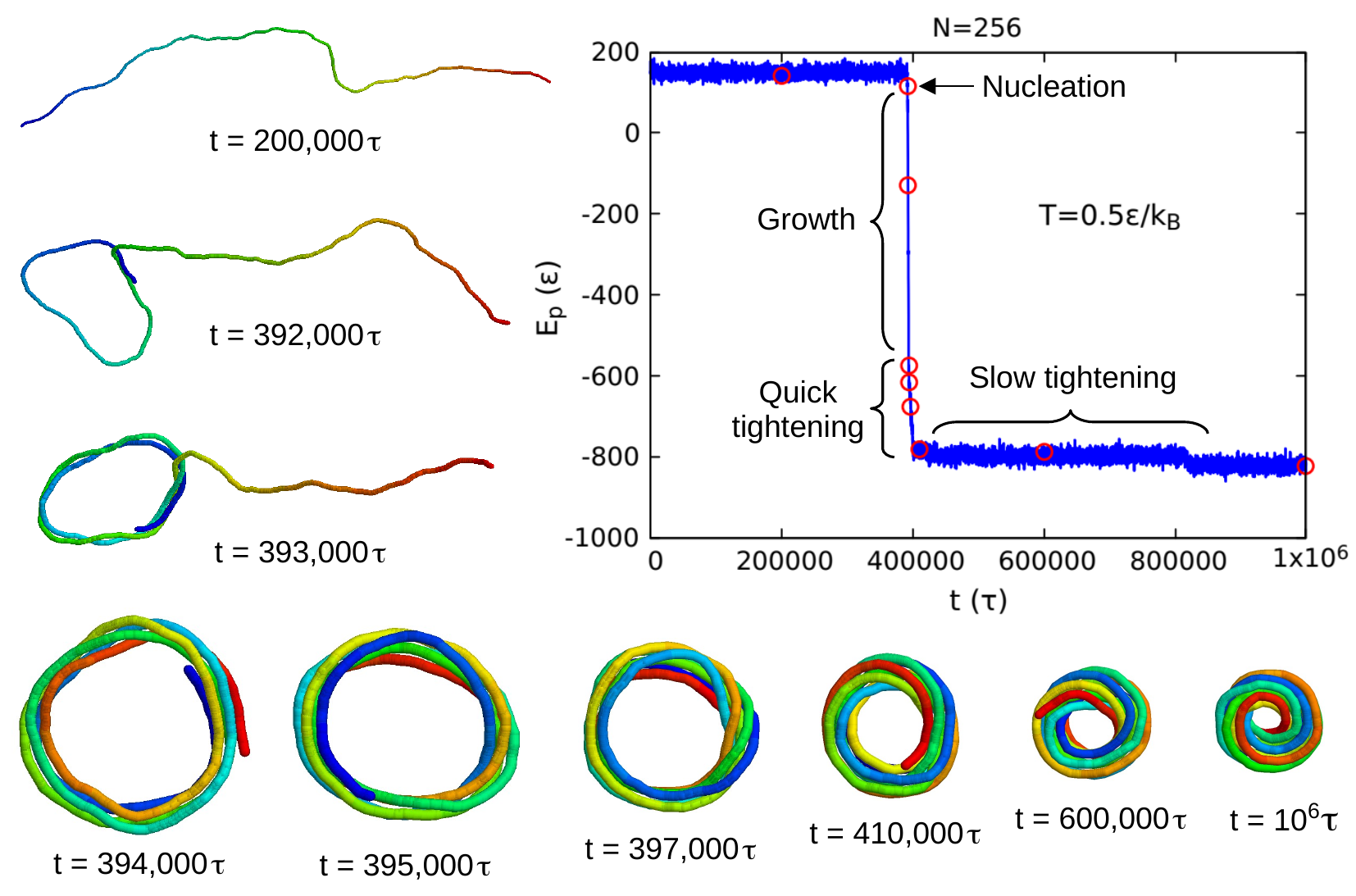}
\caption{A simulation trajectory leading to a toroid formation without
intermediates. The trajectory was obtained at temperature $T=0.5\,\epsilon/k_B$
for a stiff polymer of $N=256$ beads with the bending stiffness
$\kappa_b=22\epsilon$. The time dependence of the potential energy $E_p$ and a
number of conformations drawn from the trajectory at selected points (open
circles) are shown. The observed stages of the polymer collapse include
nucleation, growth, quick tightening and slow tightening as indicated. 
} 
\label{fig:3ctraj}
\end{figure*}

A typical direct pathway can be described by the trajectory shown in
Fig.~\ref{fig:3ctraj}, which is obtained at $T=0.5\,\epsilon/k_B$. 
In this trajectory, the extended chain spontaneously forms a loop which
nucleates a growth process, resulting in a rapid folding of the chain into a
toroid.  Once formed, the toroid continues to undergo a quick tightening
process and then a slow tightening process,
which reduce the toroid diameter and increase its thickness (see conformations
in Fig.~\ref{fig:3ctraj}). A sharp drop in energy is associated with both
the growth and the quick tightening processes, whereas the slow tightening 
decreases the energy only slightly and stops when the structure is fully
relaxed.  Interestingly, the toroid is seen to be twisted early in the
tightening processes and in the final structure, indicating that toroidal
bundles are prone to twisting with any toroid diameter.  Most of the toroids
formed with the above pathway has no U-shaped region in the final structure.
However, depending on the trajectory, the chain may also form an U-shaped
region during the growth process (see example in Fig.~S5).

\begin{figure*}
\includegraphics[width=13cm]{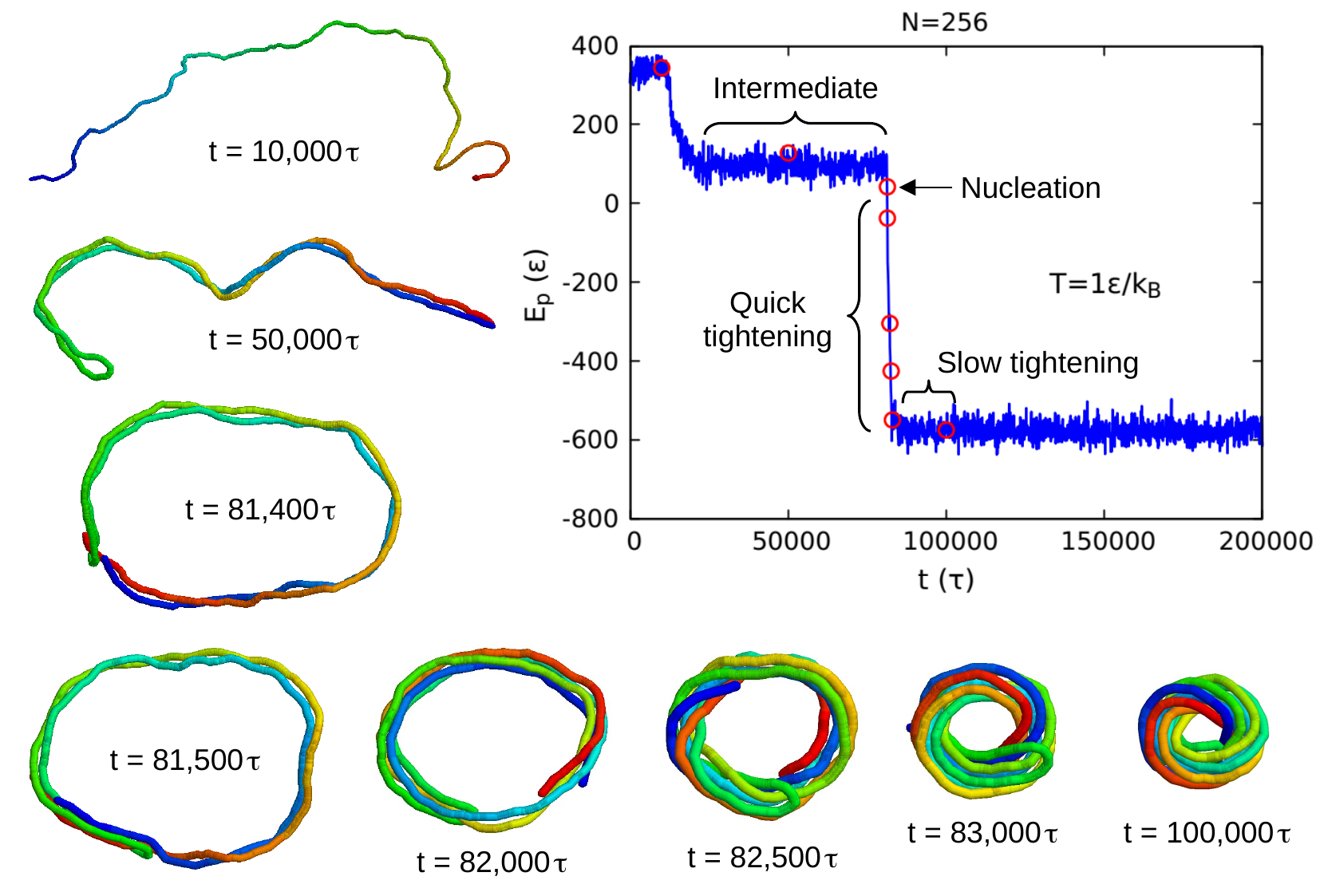}
\caption{
A simulation trajectory showing the toroid formation through a hairpin-like
intermediate. The trajectory was obtained at temperature $T=1\,\epsilon/k_B$
for the stiff polymer of $N=256$ beads with the bending stiffness
$\kappa_b=22\epsilon$. The time dependence of
the potential energy $E_p$ and several conformations drawn from the trajectory
at selected points (open circles) are shown. The stages of the polymer collapse
include an intermediate state, nucleation, quick tightening and slow tightening,
as indicated. }
\label{fig:7atraj}
\end{figure*}

An example of an indirect pathway is shown in
Fig.~\ref{fig:7atraj} with a trajectory obtained at $T=1\,\epsilon/k_B$. In this
trajectory, the chain first forms a long hairpin-like conformation (also called
one-head racquet conformation \cite{MacKintosh2004}), with the length equal
half of the DNA length, as the intermediate state. The hairpin-like
conformation then spontaneously forms a loop which then quickly tightens into a
twisted toroidal bundle (the growth process is lacking in this trajectory
because the nucleated loop has the full length of the hairpin).  The toroid
formed via this pathway typically contains the U-shaped region initially
belonging to the hairpin.  It can be expected that rod-like conformations
are possible intermediates in the toroid formation since they appear as
long-lived states in the simulations (Fig.~S4).
However, the rod to toroid transition can be extremely slow at low temperatures
due to high energy barriers. In fact, we did not observe this transition at
$T=0.5\,\epsilon/k_B$ and $T=1\,\epsilon/k_B$ within the given simulation time
limit. However, by increasing the temperature one can observe the rod to toroid
transition easily. For example, in a trajectory at $T=2\,\epsilon/k_B$ shown
Fig.~S6, the chain first forms a rod-like structure which progressively becomes
more compact and then transforms into a toroid. 
It is also found that, on average, the polymer collapse is faster as 
the temperature is increased within the low temperature range considered.
It can be understood that
increasing the temperature makes the stiff polymer more flexible and promotes
conformational changes, resulting in shorter times for nucleation loop
formation and transitions between metastable states.
Note that the change in
temperature in our model is equivalent to an inverse change in the interaction
strength between DNA segments, which can be modulated by the solvent condition.

The kinetic pathways described above are in agreement with many previous
simulation studies of semiflexible polymers
\cite{Stevens2001,Yoshikawa2002,MacKintosh2004,Muthukumar2005,Reddy2017} and are
supported by in vitro kinetic studies of DNA condensation
\cite{Hud05,Vilfan06,Yoshikawa1996jacs,Hud2000}. A plausible new finding from
the present simulations is of the tightening processes, which
are subsequent to the growth process if the nucleation loop has a large size.
Interestingly, these tightening processes dynamically facilitate the
formation of the twisted toroidal bundles. It is understood
that the twisting requires a polymer to thread through the toroid's hole many
times.  During the growth process, it is unlikely that the extended part of the
polymer can make a threading through the hole. In the tightening processes, the
threadings are possible and more easy as the polymer ends are located on the
toroid and they are pushed through the hole by the same forces that drive the
tightening. In principle, a non-twisted toroidal bundle can also
convert into a twisted bundle by a structural relaxation without the
tightening, but such a relaxation process is much slower and more difficult
due to the polymer's topological constraint. Indeed, we find that some
trajectories that lack or almost lack a tightening
process, like the one shown in Fig.~S5, only result in weakly twisted toroidal
bundles.

\section{Discussion}

The twisted bundle model represents a smooth nematic flow field. Mapping the
model to a single chain conformation requires more than a discretization. As
described before, the toroidal bundle is this model is organized into
disconnected layers of filaments. The bundle in a single layer could be
formed by a single filament, such as the one shown in Fig.~\ref{fig:tbmodel}
(b), but this may not happen in general. Connecting the filaments from
different layers lead to localized defects \cite{Kulic2004}. Thus, 
there not exists a single polymer conformation of perfectly uniform twist
vector $k$ inside the toroid. The polymer toroidal bundles obtained in our
simulations are far from the perfect one as shown by a strong variation of the
local twist vector (Fig. S3). This variation can be due to various factors
such as small system size, the chain discretization, and the surface effect etc.

It can be expected that the twisted bundles in the TB-$\rho$ model would
lead to distortion of the hexagonal lattice of filaments as the filament layers
in $\rho$ are not twisted in phase. Such a distortion would not be favorable by
the intermolecular potentials \cite{Grason2008}. Interestingly, our study
shows that the TB-$\rho$ model yields only a marginal stability of a few $k_B
T$ compared to the TB model. Thus, the uniform twisting should be more
favorable if the distortion of the hexagonal lattice in non-uniform twisting
comes with a sufficiently high energy cost.

Knotting and unknotting are rare events in polymer dynamics \cite{Micheletti2013}.
In a twisted toroidal bundle, as shown in Fig.~\ref{fig:tbmodel} (b), the DNA
chain threads through the hole of the toroid many times making the conformation
highly knotted. Given $n=\eta (\Delta/d)^2$ is the number of filament crossings
found within a toroid tubular cross section, the number of threadings
through the hole is roughly equal to $n\cdot k$ with $k$ the twist number.
As an estimate, a toroid of 100 nm diameter and with the thickness ratio
$\alpha=0.6$ would have about 31 threadings at its optimal twist number.
Such a high knotting state would be dynamically inaccessible for
a toroid formed by a single chain. However, the situation
should be much easier if the toroid is formed by multiple chains. For example,
if there are 10 chains in the toroid then each chain would have to make only
about 3 threadings.

DNA is known to have a torsional stiffness associated with elastic
responses of the molecule to twist deformations, which correspond to either
overwinding or underwinding of the DNA double helix \cite{Hagerman1988}. The
torsional stiffness plays an important role in DNA supercoiling, for example in
the wrapping of DNA around the histone protein core in nucleosomes
\cite{Kaczmarczyk2020}. Marko and Siggia \cite{Marko1994} showed that the
elasticity of DNA also includes a twist-bend coupling arising from the
asymmetry between the major and minor grooves of the double helix.  Due to this
coupling, bending the molecule induces an unwinding of the helix,
whereas overwinding it increases the helix's bending stiffness.  There is also
a softening of the bending rigidity due to the twist-bend coupling.  
In the Marko and Siggia's model of B-DNA \cite{Marko1994}, the twist strain
$\Omega_3$ is defined as a deviation of the local twist density from $(\omega_0 -
\tau_s)$, where $\omega_0=2\pi/l=1.85~\mathrm{nm}^{-1}$ is the twist density of
DNA in absence of deformations ($l=3.4$~nm is the DNA helical pitch), and
$\tau_s$ is the torsion of the molecular axis [see Eq.~(\ref{eq:FS})]. 
The torsion in the twisted bundle model can be calculated giving
$|\tau_s| < 0.04$~nm$^{-1}$
for toroids of typical sizes.
For DNA conformations with a constant curvature $c$ and a constant torsion
$\tau_s$, Marko and Siggia showed 
that the equilibrium excess twist per helix repeat is $\langle \Omega_3
\rangle/\omega_0 = - 0.5 (D/C)^2 (c/\omega_0)^2$ \cite{Marko1994} for $\tau_s=0$, 
where $C$ is the torsional stiffness and $D$ is the twist-bend coupling
constant. It was also shown that the case of $\tau_s \neq 0$ leads only to a small
correction for $\langle \Omega_3 \rangle/\omega_0$ (Fig.~2 of
Ref.~\cite{Marko1994}). 
A recent study has estimated that $C = 110~\mathrm{nm}\cdot k_BT$ and $D =
40~\mathrm{nm} \cdot k_B T$ \cite{Carlon2017}. 
For the DNA in a toroid with the mean radius $R=40$~nm, using the
constant radius of curvature approximation, we get $c = 0.025$~nm$^{-1}$, 
giving $\langle \Omega_3 \rangle \approx -2.2 \times 10^{-5}$~nm$^{-1}$.
It comes that the twist energy of the toroid, calculated as $L \frac{1}{2} C
\langle \Omega_3 \rangle^2$, is less than $0.03~k_BT$.
Thus, the contribution of the twist energy in toroids is negligible.
This result is much different from that for the DNA supercoil on the
nucleosome, for which the radius of curvature of DNA is just $\sim$4.5~nm.
With such a small radius, the twist-bend coupling can lead
to measurable effects \cite{Marko1994,Carlon2019}.

\section{Conclusion}

We have studied several theoretical models for the organization of DNA
in toroidal condensates. The results from these models show that the twisted
toroidal bundles provide the best stability for toroidal condensates.
The two models of twisted bundles considered, one with uniform and another
with non-uniform twisting, yield similar results upon energy 
minimization. These models show that a moderate twisting can substantially
lower the bending energy of the toroid. The degree of twisting
of the bundle can be quantified by the twist number $k$, which determines
how quick the change in the rotation angle $\theta$ around the toroid tubular
axis is compared to the change in the rotation angle $\phi$ around the toroid
main axis along the trace of a DNA filament in toroidal coordinates. The
uniformly twisted bundle model shows that the optimal twist number $k^*$, for
the toroid energy is minimized, ranges from 0.66 to 1.12, and depends only on
the toroid thickness ratio $\alpha$. 

Interestingly, our REMD simulations show that the ground states of stiff 
polymers with self-attractive potentials are twisted toroidal bundles, which
confirms the finding of the theoretical models. Furthermore, the average twist
numbers calculated for the twisted bundles obtained by simulations are close to
the optimal twist number $k^*$ predicted by the uniformly twisted bundle model.
Constant-temperature simulations of the stiff polymers show that toroids can be
formed through a direct pathway or via a hairpin-like or rod-like intermediate
with the collapse governed by the nucleation-growth mechanism.
The formation of twisted toroidal bundles is dynamically facilitated by the
toroid's tightening processes which happen subsequently to the growth
process if the toroid formation starts with a large nucleation loop. Due to the
topological constraint of the polymer, it is increasingly difficult for 
a polymer with an increased chain length to access the lowest energy twisted
bundle state, as found for the chain of 512 beads.
The observation of strongly twisted bundles with sharp U-shaped regions in the
simulations suggests that the problem of dynamical accessibility of the twisted
bundles can be reduced if the toroid is formed by multiple chains as found
in multimolecular DNA condensates.

It has been found that the formations of twisted bundles of filaments and
columns are often driven by molecular chirality \cite{Grason2015}.
The present study, as well as the early one \cite{Kulic2004}, indicates that
toroidal bundles are prone to twisting due to the effect of bending
stiffness, in absence of molecular chirality.
It is also shown here that folding of a polymer
into the non-trivial structures of twisted toroidal
bundles is not as difficult as it seems to be.

\section*{Supplementary Material}

See supplementary material for the curvature maps of the toroids with
thickness ratio $\alpha=0.9$; for the dependences of the optimal twist number
$k^*$ the mean radius $R$, and the thickness radius $\Delta$ on the surface
tension $\sigma$ of the toroidal condensate in the TB model; for  
the analysis of the local tangent vectors of the stiff polymers in
the toroidal bundles obtained by REMD simulations; 
for the observation of long-lived metastable states in 
constant-temperature simulations; for a simulation
trajectory showing the formation of an U-shaped region during the toroid
growth; and for a simulation trajectory with a conversion from rod to toroid.

\section*{Acknowledgements}

Nhung T. T. Nguyen was funded by Vingroup JSC and supported by the
Postdoctoral Scholarship Programme of Vingroup Innovation Foundation
(VINIF), Institute of Big Data, code VINIF.2021.STS.03.
%T.X.H. acknowledges the support of the Vietnam Academy of Science and
%Technology for high-level researchers under grant number NVCC05.05/22-23.
This research is funded by Vietnam National Foundation for Science and
Technology Development (NAFOSTED) under Grant No. 103.01-2019.363.

\section*{Data availability statement}

The data that support the findings of this study are available from the
corresponding author upon reasonable request.

\bibliography{refs_dna}

\begin{thebibliography}{51}%
\makeatletter
\providecommand \@ifxundefined [1]{%
 \@ifx{#1\undefined}
}%
\providecommand \@ifnum [1]{%
 \ifnum #1\expandafter \@firstoftwo
 \else \expandafter \@secondoftwo
 \fi
}%
\providecommand \@ifx [1]{%
 \ifx #1\expandafter \@firstoftwo
 \else \expandafter \@secondoftwo
 \fi
}%
\providecommand \natexlab [1]{#1}%
\providecommand \enquote  [1]{``#1''}%
\providecommand \bibnamefont  [1]{#1}%
\providecommand \bibfnamefont [1]{#1}%
\providecommand \citenamefont [1]{#1}%
\providecommand \href@noop [0]{\@secondoftwo}%
\providecommand \href [0]{\begingroup \@sanitize@url \@href}%
\providecommand \@href[1]{\@@startlink{#1}\@@href}%
\providecommand \@@href[1]{\endgroup#1\@@endlink}%
\providecommand \@sanitize@url [0]{\catcode `\\12\catcode `\$12\catcode
  `\&12\catcode `\#12\catcode `\^12\catcode `\_12\catcode `\%12\relax}%
\providecommand \@@startlink[1]{}%
\providecommand \@@endlink[0]{}%
\providecommand \url  [0]{\begingroup\@sanitize@url \@url }%
\providecommand \@url [1]{\endgroup\@href {#1}{\urlprefix }}%
\providecommand \urlprefix  [0]{URL }%
\providecommand \Eprint [0]{\href }%
\providecommand \doibase [0]{http://dx.doi.org/}%
\providecommand \selectlanguage [0]{\@gobble}%
\providecommand \bibinfo  [0]{\@secondoftwo}%
\providecommand \bibfield  [0]{\@secondoftwo}%
\providecommand \translation [1]{[#1]}%
\providecommand \BibitemOpen [0]{}%
\providecommand \bibitemStop [0]{}%
\providecommand \bibitemNoStop [0]{.\EOS\space}%
\providecommand \EOS [0]{\spacefactor3000\relax}%
\providecommand \BibitemShut  [1]{\csname bibitem#1\endcsname}%
\let\auto@bib@innerbib\@empty
%</preamble>
\bibitem [{\citenamefont {Bloomfield}(1996)}]{Bloomfield}%
  \BibitemOpen
  \bibfield  {author} {\bibinfo {author} {\bibfnamefont {V.~A.}\ \bibnamefont
  {Bloomfield}},\ }\href@noop {} {\bibfield  {journal} {\bibinfo  {journal}
  {Curr. Opin. Struct. Bio.}\ }\textbf {\bibinfo {volume} {6}},\ \bibinfo
  {pages} {334} (\bibinfo {year} {1996})}\BibitemShut {NoStop}%
\bibitem [{\citenamefont {Hud}\ and\ \citenamefont {Vilfan}(2005)}]{Hud05}%
  \BibitemOpen
  \bibfield  {author} {\bibinfo {author} {\bibfnamefont {N.~V.}\ \bibnamefont
  {Hud}}\ and\ \bibinfo {author} {\bibfnamefont {I.~D.}\ \bibnamefont
  {Vilfan}},\ }\href@noop {} {\bibfield  {journal} {\bibinfo  {journal} {Annu.
  Rev. Biophys. Biomol. Struct.}\ }\textbf {\bibinfo {volume} {34}},\ \bibinfo
  {pages} {295} (\bibinfo {year} {2005})}\BibitemShut {NoStop}%
\bibitem [{\citenamefont {Gosule}\ and\ \citenamefont
  {Schellman}(1976)}]{Gosule}%
  \BibitemOpen
  \bibfield  {author} {\bibinfo {author} {\bibfnamefont {L.~C.}\ \bibnamefont
  {Gosule}}\ and\ \bibinfo {author} {\bibfnamefont {J.~A.}\ \bibnamefont
  {Schellman}},\ }\href@noop {} {\bibfield  {journal} {\bibinfo  {journal}
  {Nature}\ }\textbf {\bibinfo {volume} {259}},\ \bibinfo {pages} {333}
  (\bibinfo {year} {1976})}\BibitemShut {NoStop}%
\bibitem [{\citenamefont {Strey}\ \emph {et~al.}(1998)\citenamefont {Strey},
  \citenamefont {Podgornik}, \citenamefont {Rau},\ and\ \citenamefont
  {Parsegian}}]{RP1}%
  \BibitemOpen
  \bibfield  {author} {\bibinfo {author} {\bibfnamefont {H.~H.}\ \bibnamefont
  {Strey}}, \bibinfo {author} {\bibfnamefont {R.}~\bibnamefont {Podgornik}},
  \bibinfo {author} {\bibfnamefont {D.~C.}\ \bibnamefont {Rau}}, \ and\
  \bibinfo {author} {\bibfnamefont {V.~A.}\ \bibnamefont {Parsegian}},\
  }\href@noop {} {\bibfield  {journal} {\bibinfo  {journal} {Curr. Opin.
  Struct. Bio.}\ }\textbf {\bibinfo {volume} {8}},\ \bibinfo {pages} {309}
  (\bibinfo {year} {1998})}\BibitemShut {NoStop}%
\bibitem [{\citenamefont {Maniatis}, \citenamefont {Venable~Jr},\ and\
  \citenamefont {Lerman}(1974)}]{Maniatis}%
  \BibitemOpen
  \bibfield  {author} {\bibinfo {author} {\bibfnamefont {T.}~\bibnamefont
  {Maniatis}}, \bibinfo {author} {\bibfnamefont {J.~H.}\ \bibnamefont
  {Venable~Jr}}, \ and\ \bibinfo {author} {\bibfnamefont {L.~S.}\ \bibnamefont
  {Lerman}},\ }\href@noop {} {\bibfield  {journal} {\bibinfo  {journal} {J.
  Mol. Bio.}\ }\textbf {\bibinfo {volume} {84}},\ \bibinfo {pages} {37}
  (\bibinfo {year} {1974})}\BibitemShut {NoStop}%
\bibitem [{\citenamefont {Vilfan}\ \emph {et~al.}(2006)\citenamefont {Vilfan},
  \citenamefont {Conwell}, \citenamefont {Sarkar},\ and\ \citenamefont
  {Hud}}]{Vilfan06}%
  \BibitemOpen
  \bibfield  {author} {\bibinfo {author} {\bibfnamefont {I.~D.}\ \bibnamefont
  {Vilfan}}, \bibinfo {author} {\bibfnamefont {C.~C.}\ \bibnamefont {Conwell}},
  \bibinfo {author} {\bibfnamefont {T.}~\bibnamefont {Sarkar}}, \ and\ \bibinfo
  {author} {\bibfnamefont {N.~V.}\ \bibnamefont {Hud}},\ }\href@noop {}
  {\bibfield  {journal} {\bibinfo  {journal} {Biochemistry}\ }\textbf {\bibinfo
  {volume} {45}},\ \bibinfo {pages} {8174} (\bibinfo {year}
  {2006})}\BibitemShut {NoStop}%
\bibitem [{\citenamefont {Conwell}, \citenamefont {Vilfan},\ and\ \citenamefont
  {Hud}(2003)}]{Conwell}%
  \BibitemOpen
  \bibfield  {author} {\bibinfo {author} {\bibfnamefont {C.~C.}\ \bibnamefont
  {Conwell}}, \bibinfo {author} {\bibfnamefont {I.~D.}\ \bibnamefont {Vilfan}},
  \ and\ \bibinfo {author} {\bibfnamefont {N.~V.}\ \bibnamefont {Hud}},\
  }\href@noop {} {\bibfield  {journal} {\bibinfo  {journal} {Proc. Natl. Acad.
  Sci. USA}\ }\textbf {\bibinfo {volume} {100}},\ \bibinfo {pages} {9296}
  (\bibinfo {year} {2003})}\BibitemShut {NoStop}%
\bibitem [{\citenamefont {Pinto}\ \emph {et~al.}(2009)\citenamefont {Pinto},
  \citenamefont {Mor{\'a}n}, \citenamefont {Miguel}, \citenamefont {Lindman},
  \citenamefont {Jurado},\ and\ \citenamefont {Pais}}]{Pinto}%
  \BibitemOpen
  \bibfield  {author} {\bibinfo {author} {\bibfnamefont {M.~F.}\ \bibnamefont
  {Pinto}}, \bibinfo {author} {\bibfnamefont {M.~C.}\ \bibnamefont
  {Mor{\'a}n}}, \bibinfo {author} {\bibfnamefont {M.~G.}\ \bibnamefont
  {Miguel}}, \bibinfo {author} {\bibfnamefont {B.}~\bibnamefont {Lindman}},
  \bibinfo {author} {\bibfnamefont {A.~S.}\ \bibnamefont {Jurado}}, \ and\
  \bibinfo {author} {\bibfnamefont {A.~A.}\ \bibnamefont {Pais}},\ }\href@noop
  {} {\bibfield  {journal} {\bibinfo  {journal} {Biomacromolecules}\ }\textbf
  {\bibinfo {volume} {10}},\ \bibinfo {pages} {1319} (\bibinfo {year}
  {2009})}\BibitemShut {NoStop}%
\bibitem [{\citenamefont {Arscott}, \citenamefont {Li},\ and\ \citenamefont
  {Bloomfield}(1990)}]{Bloomfield1990}%
  \BibitemOpen
  \bibfield  {author} {\bibinfo {author} {\bibfnamefont {P.~G.}\ \bibnamefont
  {Arscott}}, \bibinfo {author} {\bibfnamefont {A.-Z.}\ \bibnamefont {Li}}, \
  and\ \bibinfo {author} {\bibfnamefont {V.~A.}\ \bibnamefont {Bloomfield}},\
  }\href@noop {} {\bibfield  {journal} {\bibinfo  {journal} {Biopolymers}\
  }\textbf {\bibinfo {volume} {30}},\ \bibinfo {pages} {619} (\bibinfo {year}
  {1990})}\BibitemShut {NoStop}%
\bibitem [{\citenamefont {Grosberg}\ and\ \citenamefont
  {Zhestkov}(1986)}]{Grosberg1986}%
  \BibitemOpen
  \bibfield  {author} {\bibinfo {author} {\bibfnamefont {A.~Y.}\ \bibnamefont
  {Grosberg}}\ and\ \bibinfo {author} {\bibfnamefont {A.}~\bibnamefont
  {Zhestkov}},\ }\href@noop {} {\bibfield  {journal} {\bibinfo  {journal} {J.
  Biomol. Struct. Dyn.}\ }\textbf {\bibinfo {volume} {3}},\ \bibinfo {pages}
  {859} (\bibinfo {year} {1986})}\BibitemShut {NoStop}%
\bibitem [{\citenamefont {Vasilevskaya}\ \emph {et~al.}(1997)\citenamefont
  {Vasilevskaya}, \citenamefont {Khokhlov}, \citenamefont {Kidoaki},\ and\
  \citenamefont {Yoshikawa}}]{Vasilevskaya1997}%
  \BibitemOpen
  \bibfield  {author} {\bibinfo {author} {\bibfnamefont {V.}~\bibnamefont
  {Vasilevskaya}}, \bibinfo {author} {\bibfnamefont {A.}~\bibnamefont
  {Khokhlov}}, \bibinfo {author} {\bibfnamefont {S.}~\bibnamefont {Kidoaki}}, \
  and\ \bibinfo {author} {\bibfnamefont {K.}~\bibnamefont {Yoshikawa}},\
  }\href@noop {} {\bibfield  {journal} {\bibinfo  {journal} {Biopolymers}\
  }\textbf {\bibinfo {volume} {41}},\ \bibinfo {pages} {51} (\bibinfo {year}
  {1997})}\BibitemShut {NoStop}%
\bibitem [{\citenamefont {Noguchi}\ and\ \citenamefont
  {Yoshikawa}(1998)}]{Yoshikawa1998}%
  \BibitemOpen
  \bibfield  {author} {\bibinfo {author} {\bibfnamefont {H.}~\bibnamefont
  {Noguchi}}\ and\ \bibinfo {author} {\bibfnamefont {K.}~\bibnamefont
  {Yoshikawa}},\ }\href@noop {} {\bibfield  {journal} {\bibinfo  {journal} {J.
  Chem. Phys.}\ }\textbf {\bibinfo {volume} {109}},\ \bibinfo {pages} {5070}
  (\bibinfo {year} {1998})}\BibitemShut {NoStop}%
\bibitem [{\citenamefont {Ivanov}\ \emph {et~al.}(2000)\citenamefont {Ivanov},
  \citenamefont {Stukan}, \citenamefont {Vasilevskaya}, \citenamefont {Paul},\
  and\ \citenamefont {Binder}}]{Ivanov2000}%
  \BibitemOpen
  \bibfield  {author} {\bibinfo {author} {\bibfnamefont {V.}~\bibnamefont
  {Ivanov}}, \bibinfo {author} {\bibfnamefont {M.}~\bibnamefont {Stukan}},
  \bibinfo {author} {\bibfnamefont {V.}~\bibnamefont {Vasilevskaya}}, \bibinfo
  {author} {\bibfnamefont {W.}~\bibnamefont {Paul}}, \ and\ \bibinfo {author}
  {\bibfnamefont {K.}~\bibnamefont {Binder}},\ }\href@noop {} {\bibfield
  {journal} {\bibinfo  {journal} {Macromol. Theory Simul.}\ }\textbf {\bibinfo
  {volume} {9}},\ \bibinfo {pages} {488} (\bibinfo {year} {2000})}\BibitemShut
  {NoStop}%
\bibitem [{\citenamefont {Stukan}\ \emph {et~al.}(2003)\citenamefont {Stukan},
  \citenamefont {Ivanov}, \citenamefont {Grosberg}, \citenamefont {Paul},\ and\
  \citenamefont {Binder}}]{Stukan2003}%
  \BibitemOpen
  \bibfield  {author} {\bibinfo {author} {\bibfnamefont {M.}~\bibnamefont
  {Stukan}}, \bibinfo {author} {\bibfnamefont {V.}~\bibnamefont {Ivanov}},
  \bibinfo {author} {\bibfnamefont {A.~Y.}\ \bibnamefont {Grosberg}}, \bibinfo
  {author} {\bibfnamefont {W.}~\bibnamefont {Paul}}, \ and\ \bibinfo {author}
  {\bibfnamefont {K.}~\bibnamefont {Binder}},\ }\href@noop {} {\bibfield
  {journal} {\bibinfo  {journal} {J. Chem. Phys.}\ }\textbf {\bibinfo {volume}
  {118}},\ \bibinfo {pages} {3392} (\bibinfo {year} {2003})}\BibitemShut
  {NoStop}%
\bibitem [{\citenamefont {Hoang}\ \emph {et~al.}(2014)\citenamefont {Hoang},
  \citenamefont {Giacometti}, \citenamefont {Podgornik}, \citenamefont
  {Nguyen}, \citenamefont {Banavar},\ and\ \citenamefont
  {Maritan}}]{Hoang2014}%
  \BibitemOpen
  \bibfield  {author} {\bibinfo {author} {\bibfnamefont {T.~X.}\ \bibnamefont
  {Hoang}}, \bibinfo {author} {\bibfnamefont {A.}~\bibnamefont {Giacometti}},
  \bibinfo {author} {\bibfnamefont {R.}~\bibnamefont {Podgornik}}, \bibinfo
  {author} {\bibfnamefont {N.~T.~T.}\ \bibnamefont {Nguyen}}, \bibinfo {author}
  {\bibfnamefont {J.~R.}\ \bibnamefont {Banavar}}, \ and\ \bibinfo {author}
  {\bibfnamefont {A.}~\bibnamefont {Maritan}},\ }\href@noop {} {\bibfield
  {journal} {\bibinfo  {journal} {J. Chem. Phys.}\ }\textbf {\bibinfo {volume}
  {140}},\ \bibinfo {pages} {064902} (\bibinfo {year} {2014})}\BibitemShut
  {NoStop}%
\bibitem [{\citenamefont {Cortini}\ \emph {et~al.}(2015)\citenamefont
  {Cortini}, \citenamefont {Car{\'e}}, \citenamefont {Victor},\ and\
  \citenamefont {Barbi}}]{Barbi2015}%
  \BibitemOpen
  \bibfield  {author} {\bibinfo {author} {\bibfnamefont {R.}~\bibnamefont
  {Cortini}}, \bibinfo {author} {\bibfnamefont {B.~R.}\ \bibnamefont
  {Car{\'e}}}, \bibinfo {author} {\bibfnamefont {J.-M.}\ \bibnamefont
  {Victor}}, \ and\ \bibinfo {author} {\bibfnamefont {M.}~\bibnamefont
  {Barbi}},\ }\href@noop {} {\bibfield  {journal} {\bibinfo  {journal} {J.
  Chem. Phys.}\ }\textbf {\bibinfo {volume} {142}},\ \bibinfo {pages}
  {03B606\_1} (\bibinfo {year} {2015})}\BibitemShut {NoStop}%
\bibitem [{\citenamefont {Stukan}\ \emph {et~al.}(2006)\citenamefont {Stukan},
  \citenamefont {An}, \citenamefont {Ivanov},\ and\ \citenamefont
  {Vinogradova}}]{Stukan2006}%
  \BibitemOpen
  \bibfield  {author} {\bibinfo {author} {\bibfnamefont {M.~R.}\ \bibnamefont
  {Stukan}}, \bibinfo {author} {\bibfnamefont {E.~A.}\ \bibnamefont {An}},
  \bibinfo {author} {\bibfnamefont {V.~A.}\ \bibnamefont {Ivanov}}, \ and\
  \bibinfo {author} {\bibfnamefont {O.~I.}\ \bibnamefont {Vinogradova}},\
  }\href@noop {} {\bibfield  {journal} {\bibinfo  {journal} {Phys. Rev. E}\
  }\textbf {\bibinfo {volume} {73}},\ \bibinfo {pages} {051804} (\bibinfo
  {year} {2006})}\BibitemShut {NoStop}%
\bibitem [{\citenamefont {Hoang}\ \emph {et~al.}(2015)\citenamefont {Hoang},
  \citenamefont {Trinh}, \citenamefont {Giacometti}, \citenamefont {Podgornik},
  \citenamefont {Banavar},\ and\ \citenamefont {Maritan}}]{Hoang2015}%
  \BibitemOpen
  \bibfield  {author} {\bibinfo {author} {\bibfnamefont {T.~X.}\ \bibnamefont
  {Hoang}}, \bibinfo {author} {\bibfnamefont {H.~L.}\ \bibnamefont {Trinh}},
  \bibinfo {author} {\bibfnamefont {A.}~\bibnamefont {Giacometti}}, \bibinfo
  {author} {\bibfnamefont {R.}~\bibnamefont {Podgornik}}, \bibinfo {author}
  {\bibfnamefont {J.~R.}\ \bibnamefont {Banavar}}, \ and\ \bibinfo {author}
  {\bibfnamefont {A.}~\bibnamefont {Maritan}},\ }\href@noop {} {\bibfield
  {journal} {\bibinfo  {journal} {Phys. Rev. E}\ }\textbf {\bibinfo {volume}
  {92}},\ \bibinfo {pages} {060701(R)} (\bibinfo {year} {2015})}\BibitemShut
  {NoStop}%
\bibitem [{\citenamefont {Ishimoto}\ and\ \citenamefont
  {Kikuchi}(2008)}]{Ishimoto2008}%
  \BibitemOpen
  \bibfield  {author} {\bibinfo {author} {\bibfnamefont {Y.}~\bibnamefont
  {Ishimoto}}\ and\ \bibinfo {author} {\bibfnamefont {N.}~\bibnamefont
  {Kikuchi}},\ }\href@noop {} {\bibfield  {journal} {\bibinfo  {journal} {J.
  Chem. Phys.}\ }\textbf {\bibinfo {volume} {128}},\ \bibinfo {pages} {134906}
  (\bibinfo {year} {2008})}\BibitemShut {NoStop}%
\bibitem [{\citenamefont {Vasilevskaya}\ \emph {et~al.}(1995)\citenamefont
  {Vasilevskaya}, \citenamefont {Khokhlov}, \citenamefont {Matsuzawa},\ and\
  \citenamefont {Yoshikawa}}]{Vasilevskaya1995}%
  \BibitemOpen
  \bibfield  {author} {\bibinfo {author} {\bibfnamefont {V.}~\bibnamefont
  {Vasilevskaya}}, \bibinfo {author} {\bibfnamefont {A.}~\bibnamefont
  {Khokhlov}}, \bibinfo {author} {\bibfnamefont {Y.}~\bibnamefont {Matsuzawa}},
  \ and\ \bibinfo {author} {\bibfnamefont {K.}~\bibnamefont {Yoshikawa}},\
  }\href@noop {} {\bibfield  {journal} {\bibinfo  {journal} {J. Chem. Phys.}\
  }\textbf {\bibinfo {volume} {102}},\ \bibinfo {pages} {6595} (\bibinfo {year}
  {1995})}\BibitemShut {NoStop}%
\bibitem [{\citenamefont {Yoshikawa}\ \emph {et~al.}(1996)\citenamefont
  {Yoshikawa}, \citenamefont {Takahashi}, \citenamefont {Vasilevskaya},\ and\
  \citenamefont {Khokhlov}}]{Yoshikawa1996}%
  \BibitemOpen
  \bibfield  {author} {\bibinfo {author} {\bibfnamefont {K.}~\bibnamefont
  {Yoshikawa}}, \bibinfo {author} {\bibfnamefont {M.}~\bibnamefont
  {Takahashi}}, \bibinfo {author} {\bibfnamefont {V.}~\bibnamefont
  {Vasilevskaya}}, \ and\ \bibinfo {author} {\bibfnamefont {A.}~\bibnamefont
  {Khokhlov}},\ }\href@noop {} {\bibfield  {journal} {\bibinfo  {journal}
  {Phys. Rev. Lett.}\ }\textbf {\bibinfo {volume} {76}},\ \bibinfo {pages}
  {3029} (\bibinfo {year} {1996})}\BibitemShut {NoStop}%
\bibitem [{\citenamefont {Sakaue}\ and\ \citenamefont
  {Yoshikawa}(2002)}]{Yoshikawa2002}%
  \BibitemOpen
  \bibfield  {author} {\bibinfo {author} {\bibfnamefont {T.}~\bibnamefont
  {Sakaue}}\ and\ \bibinfo {author} {\bibfnamefont {K.}~\bibnamefont
  {Yoshikawa}},\ }\href@noop {} {\bibfield  {journal} {\bibinfo  {journal} {J.
  Chem. Phys.}\ }\textbf {\bibinfo {volume} {117}},\ \bibinfo {pages} {6323}
  (\bibinfo {year} {2002})}\BibitemShut {NoStop}%
\bibitem [{\citenamefont {Montesi}, \citenamefont {Pasquali},\ and\
  \citenamefont {MacKintosh}(2004)}]{MacKintosh2004}%
  \BibitemOpen
  \bibfield  {author} {\bibinfo {author} {\bibfnamefont {A.}~\bibnamefont
  {Montesi}}, \bibinfo {author} {\bibfnamefont {M.}~\bibnamefont {Pasquali}}, \
  and\ \bibinfo {author} {\bibfnamefont {F.}~\bibnamefont {MacKintosh}},\
  }\href@noop {} {\bibfield  {journal} {\bibinfo  {journal} {Phys. Rev. E}\
  }\textbf {\bibinfo {volume} {69}},\ \bibinfo {pages} {021916} (\bibinfo
  {year} {2004})}\BibitemShut {NoStop}%
\bibitem [{\citenamefont {Ou}\ and\ \citenamefont
  {Muthukumar}(2005)}]{Muthukumar2005}%
  \BibitemOpen
  \bibfield  {author} {\bibinfo {author} {\bibfnamefont {Z.}~\bibnamefont
  {Ou}}\ and\ \bibinfo {author} {\bibfnamefont {M.}~\bibnamefont
  {Muthukumar}},\ }\href@noop {} {\bibfield  {journal} {\bibinfo  {journal} {J.
  Chem. Phys.}\ }\textbf {\bibinfo {volume} {123}},\ \bibinfo {pages} {074905}
  (\bibinfo {year} {2005})}\BibitemShut {NoStop}%
\bibitem [{\citenamefont {Dey}\ and\ \citenamefont {Reddy}(2017)}]{Reddy2017}%
  \BibitemOpen
  \bibfield  {author} {\bibinfo {author} {\bibfnamefont {A.}~\bibnamefont
  {Dey}}\ and\ \bibinfo {author} {\bibfnamefont {G.}~\bibnamefont {Reddy}},\
  }\href@noop {} {\bibfield  {journal} {\bibinfo  {journal} {J. Phys. Chem. B}\
  }\textbf {\bibinfo {volume} {121}},\ \bibinfo {pages} {9291} (\bibinfo {year}
  {2017})}\BibitemShut {NoStop}%
\bibitem [{\citenamefont {Schellman}\ and\ \citenamefont
  {Parthasarathy}(1984)}]{Schellman84}%
  \BibitemOpen
  \bibfield  {author} {\bibinfo {author} {\bibfnamefont {J.~A.}\ \bibnamefont
  {Schellman}}\ and\ \bibinfo {author} {\bibfnamefont {N.}~\bibnamefont
  {Parthasarathy}},\ }\href@noop {} {\bibfield  {journal} {\bibinfo  {journal}
  {J. Mol. Bio.}\ }\textbf {\bibinfo {volume} {175}},\ \bibinfo {pages} {313}
  (\bibinfo {year} {1984})}\BibitemShut {NoStop}%
\bibitem [{\citenamefont {Hud}\ and\ \citenamefont {Downing}(2001)}]{Hud2001}%
  \BibitemOpen
  \bibfield  {author} {\bibinfo {author} {\bibfnamefont {N.~V.}\ \bibnamefont
  {Hud}}\ and\ \bibinfo {author} {\bibfnamefont {K.~H.}\ \bibnamefont
  {Downing}},\ }\href@noop {} {\bibfield  {journal} {\bibinfo  {journal} {Proc.
  Natl. Acad. Sci. USA}\ }\textbf {\bibinfo {volume} {98}},\ \bibinfo {pages}
  {14925} (\bibinfo {year} {2001})}\BibitemShut {NoStop}%
\bibitem [{\citenamefont {Leforestier}\ and\ \citenamefont
  {Livolant}(2009)}]{Livolant2009}%
  \BibitemOpen
  \bibfield  {author} {\bibinfo {author} {\bibfnamefont {A.}~\bibnamefont
  {Leforestier}}\ and\ \bibinfo {author} {\bibfnamefont {F.}~\bibnamefont
  {Livolant}},\ }\href@noop {} {\bibfield  {journal} {\bibinfo  {journal}
  {Proc. Natl. Acad. Sci. USA}\ }\textbf {\bibinfo {volume} {106}},\ \bibinfo
  {pages} {9157} (\bibinfo {year} {2009})}\BibitemShut {NoStop}%
\bibitem [{\citenamefont {Marx}\ and\ \citenamefont {Ruben}(1983)}]{Marx1983}%
  \BibitemOpen
  \bibfield  {author} {\bibinfo {author} {\bibfnamefont {K.~A.}\ \bibnamefont
  {Marx}}\ and\ \bibinfo {author} {\bibfnamefont {G.~C.}\ \bibnamefont
  {Ruben}},\ }\href@noop {} {\bibfield  {journal} {\bibinfo  {journal} {Nucleic
  Acids Res.}\ }\textbf {\bibinfo {volume} {11}},\ \bibinfo {pages} {1839}
  (\bibinfo {year} {1983})}\BibitemShut {NoStop}%
\bibitem [{\citenamefont {Hud}, \citenamefont {Downing},\ and\ \citenamefont
  {Balhorn}(1995)}]{Hud1995}%
  \BibitemOpen
  \bibfield  {author} {\bibinfo {author} {\bibfnamefont {N.~V.}\ \bibnamefont
  {Hud}}, \bibinfo {author} {\bibfnamefont {K.~H.}\ \bibnamefont {Downing}}, \
  and\ \bibinfo {author} {\bibfnamefont {R.}~\bibnamefont {Balhorn}},\
  }\href@noop {} {\bibfield  {journal} {\bibinfo  {journal} {Proc. Natl. Acad.
  Sci. USA}\ }\textbf {\bibinfo {volume} {92}},\ \bibinfo {pages} {3581}
  (\bibinfo {year} {1995})}\BibitemShut {NoStop}%
\bibitem [{\citenamefont {Stevens}(2001)}]{Stevens2001}%
  \BibitemOpen
  \bibfield  {author} {\bibinfo {author} {\bibfnamefont {M.~J.}\ \bibnamefont
  {Stevens}},\ }\href@noop {} {\bibfield  {journal} {\bibinfo  {journal}
  {Biophys. J.}\ }\textbf {\bibinfo {volume} {80}},\ \bibinfo {pages} {130}
  (\bibinfo {year} {2001})}\BibitemShut {NoStop}%
\bibitem [{\citenamefont {Arsuaga}\ \emph {et~al.}(2002)\citenamefont
  {Arsuaga}, \citenamefont {V{\'a}zquez}, \citenamefont {Trigueros},
  \citenamefont {Sumners},\ and\ \citenamefont {Roca}}]{Roca2002}%
  \BibitemOpen
  \bibfield  {author} {\bibinfo {author} {\bibfnamefont {J.}~\bibnamefont
  {Arsuaga}}, \bibinfo {author} {\bibfnamefont {M.}~\bibnamefont
  {V{\'a}zquez}}, \bibinfo {author} {\bibfnamefont {S.}~\bibnamefont
  {Trigueros}}, \bibinfo {author} {\bibfnamefont {D.~W.}\ \bibnamefont
  {Sumners}}, \ and\ \bibinfo {author} {\bibfnamefont {J.}~\bibnamefont
  {Roca}},\ }\href@noop {} {\bibfield  {journal} {\bibinfo  {journal} {Proc.
  Natl. Acad. Sci. USA}\ }\textbf {\bibinfo {volume} {99}},\ \bibinfo {pages}
  {5373} (\bibinfo {year} {2002})}\BibitemShut {NoStop}%
\bibitem [{\citenamefont {Kuli{\'c}}, \citenamefont {Andrienko},\ and\
  \citenamefont {Deserno}(2004)}]{Kulic2004}%
  \BibitemOpen
  \bibfield  {author} {\bibinfo {author} {\bibfnamefont {I.}~\bibnamefont
  {Kuli{\'c}}}, \bibinfo {author} {\bibfnamefont {D.}~\bibnamefont
  {Andrienko}}, \ and\ \bibinfo {author} {\bibfnamefont {M.}~\bibnamefont
  {Deserno}},\ }\href@noop {} {\bibfield  {journal} {\bibinfo  {journal} {EPL
  (Europhysics Letters)}\ }\textbf {\bibinfo {volume} {67}},\ \bibinfo {pages}
  {418} (\bibinfo {year} {2004})}\BibitemShut {NoStop}%
\bibitem [{\citenamefont {Grason}(2015)}]{Grason2015}%
  \BibitemOpen
  \bibfield  {author} {\bibinfo {author} {\bibfnamefont {G.~M.}\ \bibnamefont
  {Grason}},\ }\href@noop {} {\bibfield  {journal} {\bibinfo  {journal} {Rev.
  Mod. Phys.}\ }\textbf {\bibinfo {volume} {87}},\ \bibinfo {pages} {401}
  (\bibinfo {year} {2015})}\BibitemShut {NoStop}%
\bibitem [{\citenamefont {Sugita}\ and\ \citenamefont
  {Okamoto}(1999)}]{Sugita1999}%
  \BibitemOpen
  \bibfield  {author} {\bibinfo {author} {\bibfnamefont {Y.}~\bibnamefont
  {Sugita}}\ and\ \bibinfo {author} {\bibfnamefont {Y.}~\bibnamefont
  {Okamoto}},\ }\href@noop {} {\bibfield  {journal} {\bibinfo  {journal} {Chem.
  Phys. Lett.}\ }\textbf {\bibinfo {volume} {314}},\ \bibinfo {pages} {141}
  (\bibinfo {year} {1999})}\BibitemShut {NoStop}%
\bibitem [{\citenamefont {Kreyszig}(1991)}]{Kreyszig}%
  \BibitemOpen
  \bibfield  {author} {\bibinfo {author} {\bibfnamefont {E.}~\bibnamefont
  {Kreyszig}},\ }\href@noop {} {\emph {\bibinfo {title} {Differential
  geometry}}}\ (\bibinfo  {publisher} {Dover},\ \bibinfo {address} {New York},\
  \bibinfo {year} {1991})\BibitemShut {NoStop}%
\bibitem [{\citenamefont {Marko}\ and\ \citenamefont
  {Siggia}(1995)}]{Marko1995}%
  \BibitemOpen
  \bibfield  {author} {\bibinfo {author} {\bibfnamefont {J.~F.}\ \bibnamefont
  {Marko}}\ and\ \bibinfo {author} {\bibfnamefont {E.~D.}\ \bibnamefont
  {Siggia}},\ }\href@noop {} {\bibfield  {journal} {\bibinfo  {journal}
  {Macromolecules}\ }\textbf {\bibinfo {volume} {28}},\ \bibinfo {pages} {8759}
  (\bibinfo {year} {1995})}\BibitemShut {NoStop}%
\bibitem [{\citenamefont {Barberi}\ \emph {et~al.}(2021)\citenamefont
  {Barberi}, \citenamefont {Livolant}, \citenamefont {Leforestier},\ and\
  \citenamefont {Lenz}}]{Barberi2021}%
  \BibitemOpen
  \bibfield  {author} {\bibinfo {author} {\bibfnamefont {L.}~\bibnamefont
  {Barberi}}, \bibinfo {author} {\bibfnamefont {F.}~\bibnamefont {Livolant}},
  \bibinfo {author} {\bibfnamefont {A.}~\bibnamefont {Leforestier}}, \ and\
  \bibinfo {author} {\bibfnamefont {M.}~\bibnamefont {Lenz}},\ }\href@noop {}
  {\bibfield  {journal} {\bibinfo  {journal} {Nucleic Acids Res.}\ }\textbf
  {\bibinfo {volume} {49}},\ \bibinfo {pages} {3709} (\bibinfo {year}
  {2021})}\BibitemShut {NoStop}%
\bibitem [{\citenamefont {Marko}\ and\ \citenamefont
  {Siggia}(1994)}]{Marko1994}%
  \BibitemOpen
  \bibfield  {author} {\bibinfo {author} {\bibfnamefont {J.~F.}\ \bibnamefont
  {Marko}}\ and\ \bibinfo {author} {\bibfnamefont {E.~D.}\ \bibnamefont
  {Siggia}},\ }\href@noop {} {\bibfield  {journal} {\bibinfo  {journal}
  {Macromolecules}\ }\textbf {\bibinfo {volume} {27}},\ \bibinfo {pages} {981}
  (\bibinfo {year} {1994})}\BibitemShut {NoStop}%
\bibitem [{\citenamefont {Rosa}\ \emph {et~al.}(2003)\citenamefont {Rosa},
  \citenamefont {Hoang}, \citenamefont {Marenduzzo},\ and\ \citenamefont
  {Maritan}}]{Rosa2003}%
  \BibitemOpen
  \bibfield  {author} {\bibinfo {author} {\bibfnamefont {A.}~\bibnamefont
  {Rosa}}, \bibinfo {author} {\bibfnamefont {T.~X.}\ \bibnamefont {Hoang}},
  \bibinfo {author} {\bibfnamefont {D.}~\bibnamefont {Marenduzzo}}, \ and\
  \bibinfo {author} {\bibfnamefont {A.}~\bibnamefont {Maritan}},\ }\href@noop
  {} {\bibfield  {journal} {\bibinfo  {journal} {Macromolecules}\ }\textbf
  {\bibinfo {volume} {36}},\ \bibinfo {pages} {10095} (\bibinfo {year}
  {2003})}\BibitemShut {NoStop}%
\bibitem [{\citenamefont {Todd}\ \emph {et~al.}(2008)\citenamefont {Todd},
  \citenamefont {Parsegian}, \citenamefont {Shirahata}, \citenamefont
  {Thomas},\ and\ \citenamefont {Rau}}]{Todd}%
  \BibitemOpen
  \bibfield  {author} {\bibinfo {author} {\bibfnamefont {B.~A.}\ \bibnamefont
  {Todd}}, \bibinfo {author} {\bibfnamefont {V.~A.}\ \bibnamefont {Parsegian}},
  \bibinfo {author} {\bibfnamefont {A.}~\bibnamefont {Shirahata}}, \bibinfo
  {author} {\bibfnamefont {T.}~\bibnamefont {Thomas}}, \ and\ \bibinfo {author}
  {\bibfnamefont {D.~C.}\ \bibnamefont {Rau}},\ }\href@noop {} {\bibfield
  {journal} {\bibinfo  {journal} {Biophys. J.}\ }\textbf {\bibinfo {volume}
  {94}},\ \bibinfo {pages} {4775} (\bibinfo {year} {2008})}\BibitemShut
  {NoStop}%
\bibitem [{\citenamefont {Swendsen}\ and\ \citenamefont
  {Wang}(1986)}]{Swendsen1986}%
  \BibitemOpen
  \bibfield  {author} {\bibinfo {author} {\bibfnamefont {R.~H.}\ \bibnamefont
  {Swendsen}}\ and\ \bibinfo {author} {\bibfnamefont {J.-S.}\ \bibnamefont
  {Wang}},\ }\href@noop {} {\bibfield  {journal} {\bibinfo  {journal} {Phys.
  Rev. Lett.}\ }\textbf {\bibinfo {volume} {57}},\ \bibinfo {pages} {2607}
  (\bibinfo {year} {1986})}\BibitemShut {NoStop}%
\bibitem [{\citenamefont {Bui}\ and\ \citenamefont {Hoang}(2016)}]{Thuy2016}%
  \BibitemOpen
  \bibfield  {author} {\bibinfo {author} {\bibfnamefont {P.~T.}\ \bibnamefont
  {Bui}}\ and\ \bibinfo {author} {\bibfnamefont {T.~X.}\ \bibnamefont
  {Hoang}},\ }\href@noop {} {\bibfield  {journal} {\bibinfo  {journal} {J.
  Chem. Phys.}\ }\textbf {\bibinfo {volume} {144}},\ \bibinfo {pages} {095102}
  (\bibinfo {year} {2016})}\BibitemShut {NoStop}%
\bibitem [{\citenamefont {Yoshikawa}\ and\ \citenamefont
  {Matsuzawa}(1996)}]{Yoshikawa1996jacs}%
  \BibitemOpen
  \bibfield  {author} {\bibinfo {author} {\bibfnamefont {K.}~\bibnamefont
  {Yoshikawa}}\ and\ \bibinfo {author} {\bibfnamefont {Y.}~\bibnamefont
  {Matsuzawa}},\ }\href@noop {} {\bibfield  {journal} {\bibinfo  {journal} {J.
  Am. Chem. Soc.}\ }\textbf {\bibinfo {volume} {118}},\ \bibinfo {pages} {929}
  (\bibinfo {year} {1996})}\BibitemShut {NoStop}%
\bibitem [{\citenamefont {Shen}\ \emph {et~al.}(2000)\citenamefont {Shen},
  \citenamefont {Downing}, \citenamefont {Balhorn},\ and\ \citenamefont
  {Hud}}]{Hud2000}%
  \BibitemOpen
  \bibfield  {author} {\bibinfo {author} {\bibfnamefont {M.~R.}\ \bibnamefont
  {Shen}}, \bibinfo {author} {\bibfnamefont {K.~H.}\ \bibnamefont {Downing}},
  \bibinfo {author} {\bibfnamefont {R.}~\bibnamefont {Balhorn}}, \ and\
  \bibinfo {author} {\bibfnamefont {N.~V.}\ \bibnamefont {Hud}},\ }\href@noop
  {} {\bibfield  {journal} {\bibinfo  {journal} {J. Am. Chem. Soc.}\ }\textbf
  {\bibinfo {volume} {122}},\ \bibinfo {pages} {4833} (\bibinfo {year}
  {2000})}\BibitemShut {NoStop}%
\bibitem [{\citenamefont {Grason}(2008)}]{Grason2008}%
  \BibitemOpen
  \bibfield  {author} {\bibinfo {author} {\bibfnamefont {G.~M.}\ \bibnamefont
  {Grason}},\ }\href@noop {} {\bibfield  {journal} {\bibinfo  {journal} {EPL
  (Europhysics Letters)}\ }\textbf {\bibinfo {volume} {83}},\ \bibinfo {pages}
  {58003} (\bibinfo {year} {2008})}\BibitemShut {NoStop}%
\bibitem [{\citenamefont {Tubiana}\ \emph {et~al.}(2013)\citenamefont
  {Tubiana}, \citenamefont {Rosa}, \citenamefont {Fragiacomo},\ and\
  \citenamefont {Micheletti}}]{Micheletti2013}%
  \BibitemOpen
  \bibfield  {author} {\bibinfo {author} {\bibfnamefont {L.}~\bibnamefont
  {Tubiana}}, \bibinfo {author} {\bibfnamefont {A.}~\bibnamefont {Rosa}},
  \bibinfo {author} {\bibfnamefont {F.}~\bibnamefont {Fragiacomo}}, \ and\
  \bibinfo {author} {\bibfnamefont {C.}~\bibnamefont {Micheletti}},\
  }\href@noop {} {\bibfield  {journal} {\bibinfo  {journal} {Macromolecules}\
  }\textbf {\bibinfo {volume} {46}},\ \bibinfo {pages} {3669} (\bibinfo {year}
  {2013})}\BibitemShut {NoStop}%
\bibitem [{\citenamefont {Hagerman}(1988)}]{Hagerman1988}%
  \BibitemOpen
  \bibfield  {author} {\bibinfo {author} {\bibfnamefont {P.~J.}\ \bibnamefont
  {Hagerman}},\ }\href@noop {} {\bibfield  {journal} {\bibinfo  {journal} {Ann.
  Rev. Biophys. Biophys. Chem.}\ }\textbf {\bibinfo {volume} {17}},\ \bibinfo
  {pages} {265} (\bibinfo {year} {1988})}\BibitemShut {NoStop}%
\bibitem [{\citenamefont {Kaczmarczyk}\ \emph {et~al.}(2020)\citenamefont
  {Kaczmarczyk}, \citenamefont {Meng}, \citenamefont {Ordu}, \citenamefont
  {Noort},\ and\ \citenamefont {Dekker}}]{Kaczmarczyk2020}%
  \BibitemOpen
  \bibfield  {author} {\bibinfo {author} {\bibfnamefont {A.}~\bibnamefont
  {Kaczmarczyk}}, \bibinfo {author} {\bibfnamefont {H.}~\bibnamefont {Meng}},
  \bibinfo {author} {\bibfnamefont {O.}~\bibnamefont {Ordu}}, \bibinfo {author}
  {\bibfnamefont {J.~v.}\ \bibnamefont {Noort}}, \ and\ \bibinfo {author}
  {\bibfnamefont {N.~H.}\ \bibnamefont {Dekker}},\ }\href@noop {} {\bibfield
  {journal} {\bibinfo  {journal} {Nat. Commun.}\ }\textbf {\bibinfo {volume}
  {11}},\ \bibinfo {pages} {1} (\bibinfo {year} {2020})}\BibitemShut {NoStop}%
\bibitem [{\citenamefont {Nomidis}\ \emph {et~al.}(2017)\citenamefont
  {Nomidis}, \citenamefont {Kriegel}, \citenamefont {Vanderlinden},
  \citenamefont {Lipfert},\ and\ \citenamefont {Carlon}}]{Carlon2017}%
  \BibitemOpen
  \bibfield  {author} {\bibinfo {author} {\bibfnamefont {S.~K.}\ \bibnamefont
  {Nomidis}}, \bibinfo {author} {\bibfnamefont {F.}~\bibnamefont {Kriegel}},
  \bibinfo {author} {\bibfnamefont {W.}~\bibnamefont {Vanderlinden}}, \bibinfo
  {author} {\bibfnamefont {J.}~\bibnamefont {Lipfert}}, \ and\ \bibinfo
  {author} {\bibfnamefont {E.}~\bibnamefont {Carlon}},\ }\href@noop {}
  {\bibfield  {journal} {\bibinfo  {journal} {Phys. Rev. Lett.}\ }\textbf
  {\bibinfo {volume} {118}},\ \bibinfo {pages} {217801} (\bibinfo {year}
  {2017})}\BibitemShut {NoStop}%
\bibitem [{\citenamefont {Skoruppa}\ \emph {et~al.}(2018)\citenamefont
  {Skoruppa}, \citenamefont {Nomidis}, \citenamefont {Marko},\ and\
  \citenamefont {Carlon}}]{Carlon2019}%
  \BibitemOpen
  \bibfield  {author} {\bibinfo {author} {\bibfnamefont {E.}~\bibnamefont
  {Skoruppa}}, \bibinfo {author} {\bibfnamefont {S.~K.}\ \bibnamefont
  {Nomidis}}, \bibinfo {author} {\bibfnamefont {J.~F.}\ \bibnamefont {Marko}},
  \ and\ \bibinfo {author} {\bibfnamefont {E.}~\bibnamefont {Carlon}},\
  }\href@noop {} {\bibfield  {journal} {\bibinfo  {journal} {Phys. Rev. Lett.}\
  }\textbf {\bibinfo {volume} {121}},\ \bibinfo {pages} {088101} (\bibinfo
  {year} {2018})}\BibitemShut {NoStop}%
\end{thebibliography}%

\clearpage

\setcounter{equation}{0}
\renewcommand\theequation{S\arabic{equation}}

\setcounter{figure}{0}
\renewcommand\thefigure{S\arabic{figure}}

\setcounter{table}{0}
\renewcommand\thetable{S\arabic{table}}

\setcounter{page}{1}
\renewcommand{\bibnumfmt}[1]{[S#1]}
\renewcommand{\citenumfont}[1]{[S#1]}

\onecolumngrid

\makeatletter

\centerline{\large\bf Supplementary Material}

\begin{center}
{\large\bf Energetic preference and topological constraint effects on the
formation of DNA twisted toroidal bundles}
\end{center}

\centerline{N. T. T. Nguyen$^1$, A. T. Ngo$^2$, T. X. Hoang$^{1,3}$}

\begin{center}
{\it $^1$Institute of Physics, Vietnam Academy of Science and Technology, 10 Dao Tan,
Ba Dinh, Hanoi 11108, Vietnam\\
$^2$Chemical Engineering Department, University of Illinois at Chicago, Chicago,
IL, 60608, USA\\
$^3$Graduate University of Science and Technology, Vietnam Academy of Science
and Technology, 18 Hoang Quoc Viet, Cau Giay, Hanoi 11307, Vietnam}
\end{center}

\begin{figure}[!ht]
\includegraphics[width=8.5cm]{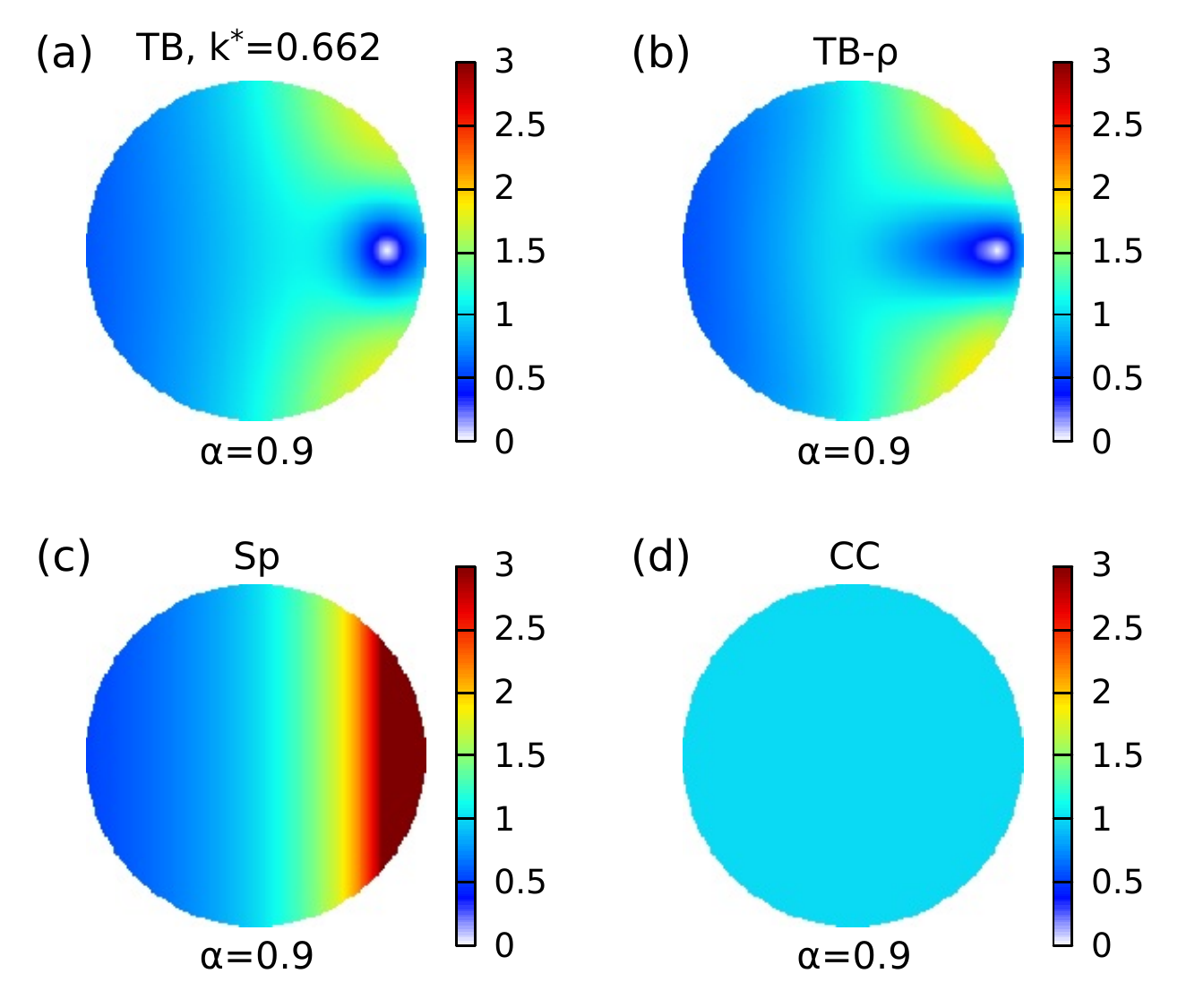}
\caption{Curvatures of DNA in modelled toroids with thickness ratio
$\alpha=0.9$. The curvatures, with values indicated by the color bar in units
of $R^{-1}$, are shown as color maps for a tubular cross section of the
toroidal bundles in the TB
(a), TB-$\rho$ (b), Sp
(c), and CC (d) models. The toroidal bundle in the TB model has the twist
number $k^*=0.662$. The right edge of the cross section corresponds to 
the inner edge of the toroid.
} \label{fig:cmap09}
\end{figure}

\vspace{20pt}
\begin{figure}[!ht]
\includegraphics[width=8.5cm]{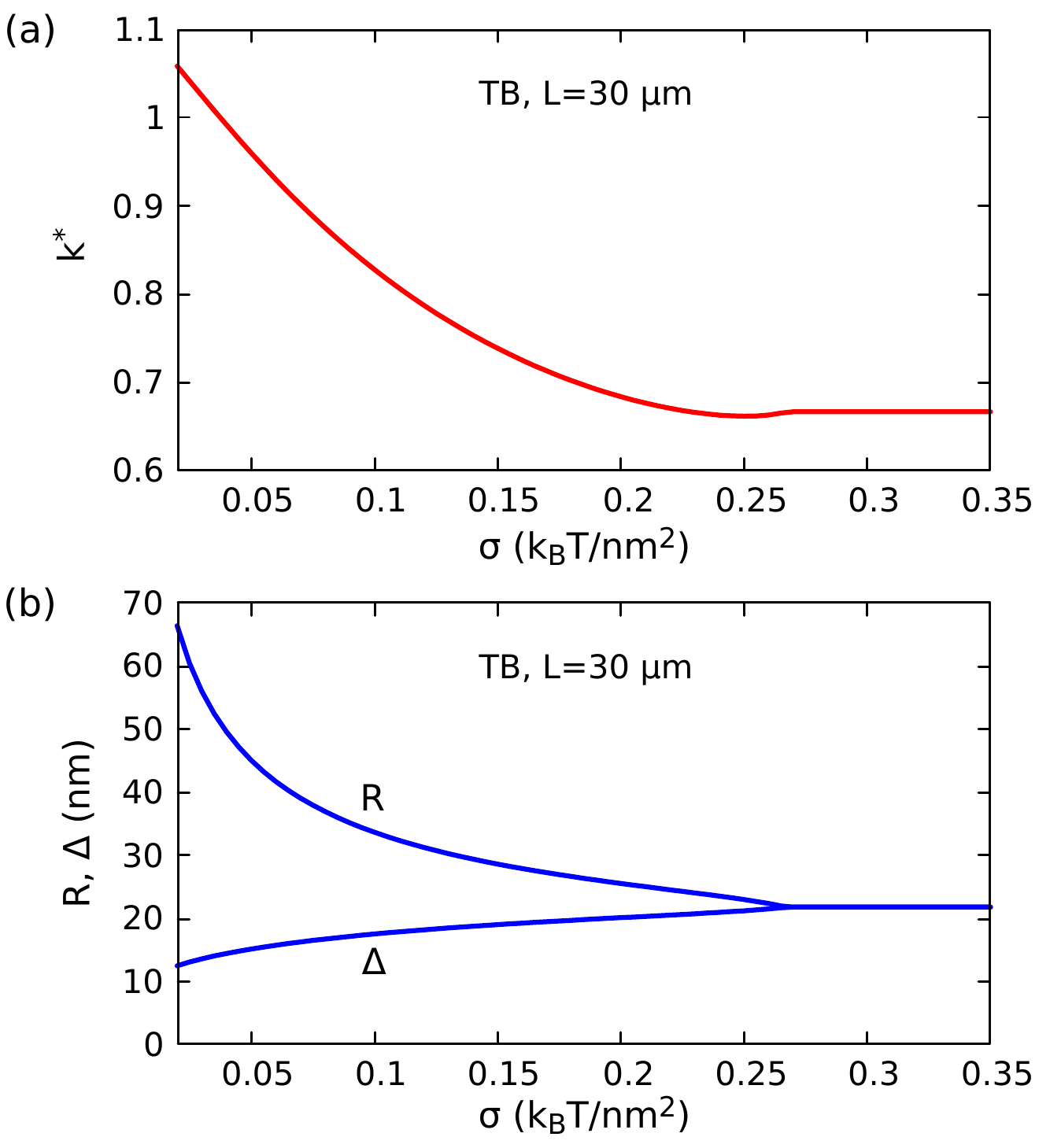}
\caption{Dependence of the optimal twist number, $k^*$, (a), the
mean radius, $R$, (b) and the thickness radius, $\Delta$, (b) on the surface
tension, $\sigma$, of the toroidal condensate in the twisted bundle model.
The data are shown for DNA length $L=30~\mu$m.
}
\label{fig:ksig}
\end{figure}

\begin{figure}[!ht]
\center
\includegraphics[width=15cm]{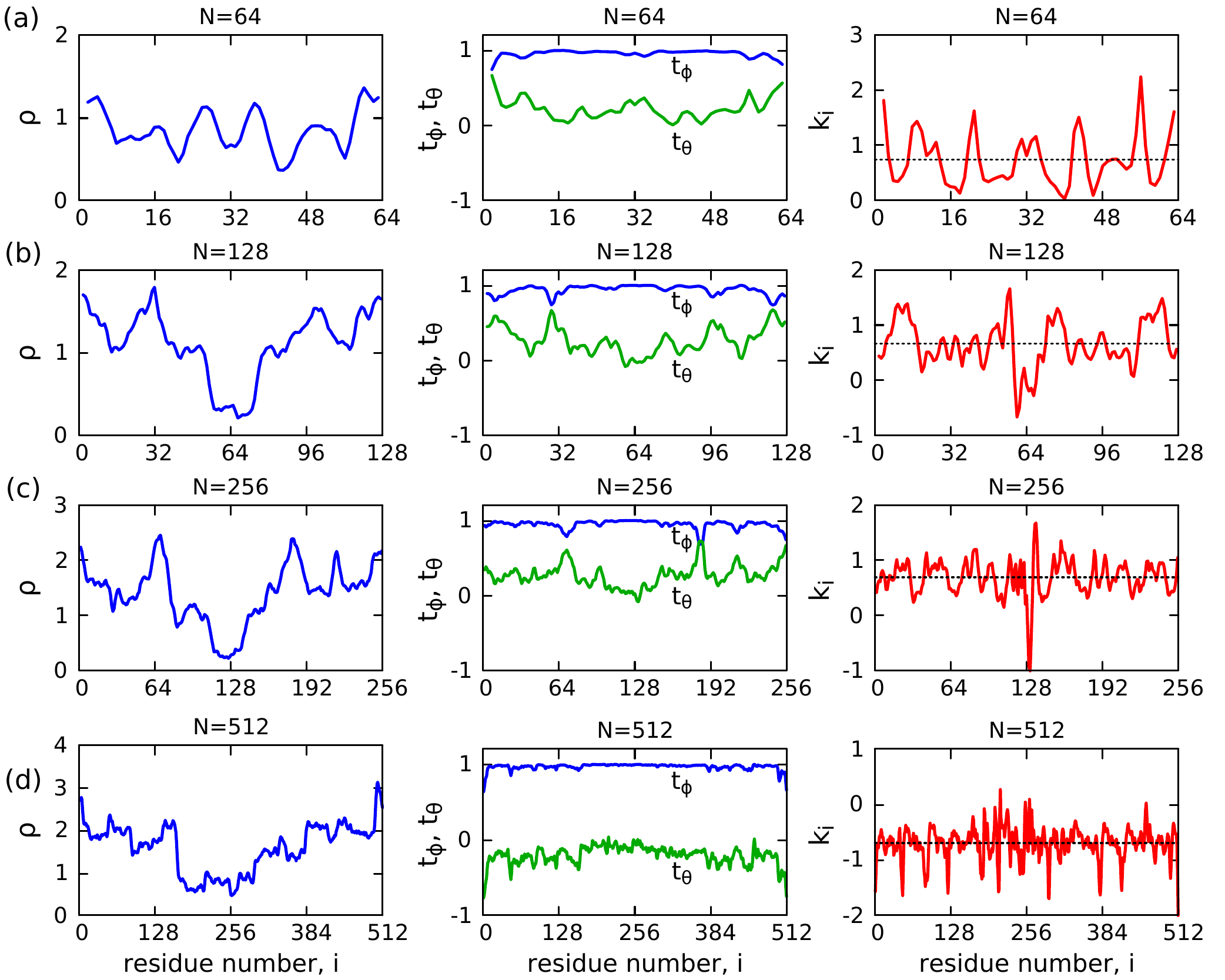} 
\caption{Dependences of the radial distance $\rho$ from the tubular axis (left
panels), and of the $\phi$- and $\theta$-components, $t_{\phi}$ and
$t_{\theta}$, of the local tangent vector ${\bf t}_i$ (middle panels) and the
local twist number $k_i$ (right panels) on the residue number, $i$, along the
chain of the stiff polymers in the toroidal conformations shown in Fig. 6
of the main text. The polymers are identified by their lengths, $N$, which are
equal to 64 (a), 128 (b), 256 (c), and 512 (d) as indicated. 
The horizontal dotted lines of the right panels indicate the mean values
$\langle k \rangle$ of the local twist numbers, which are approximately equal
to  0.73, 0.67, 0.69 and 0.7, for $N=64$, 128, 256, and 512, respectively.
} \label{fig:ktor}
\end{figure}

\begin{figure}
\includegraphics[width=14cm]{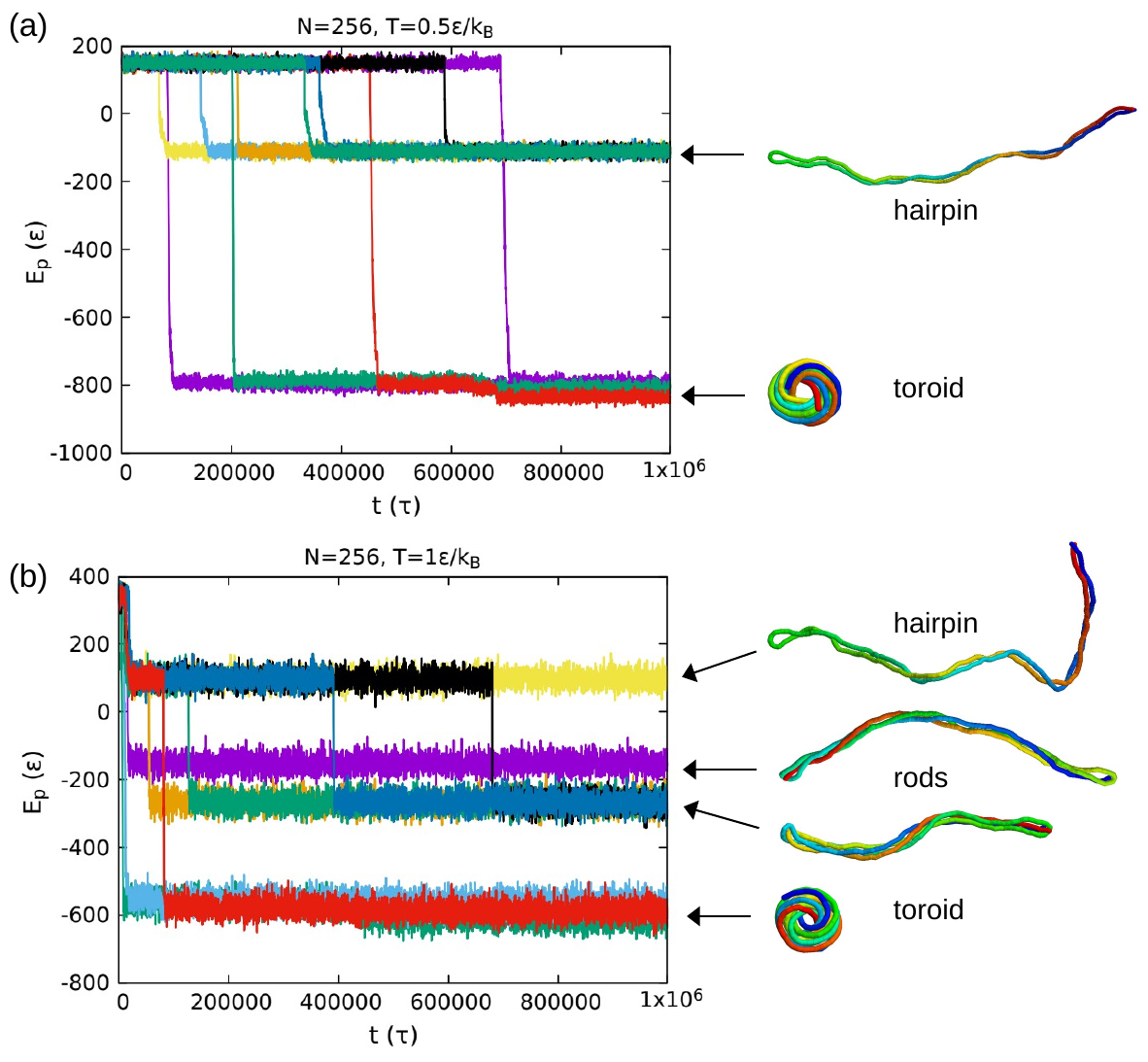}
\caption{The time dependence of the potential energy $E_p$ in some
trajectories obtained by constant-temperature simulations of stiff polymers.
The data are obtained for the stiff polymer of $N=256$ beads with the bending
stiffness $\kappa_b=22\epsilon$ at temperatures $T=0.5\,\epsilon/k_B$ (a) and
$T=1\,\epsilon/k_B$ (b). All trajectories start from a random coil
conformation. For each temperature, we
have run 32 independent trajectories but for clarity only 10 trajectories are
shown in this figure with different colors. The plateaux in $E_p$ correspond to
metastable states in the forms of hairpins and rods, and the plausible ground
state in the toroid form. Examples of conformations belonging to these states
are shown on the right. Note that some transitions from hairpin to rod and 
from hairpin to toroid are observed in the trajectories shown in panel (b).}
\label{fig:alltraj}
\end{figure}

\begin{figure}
\includegraphics[width=14cm]{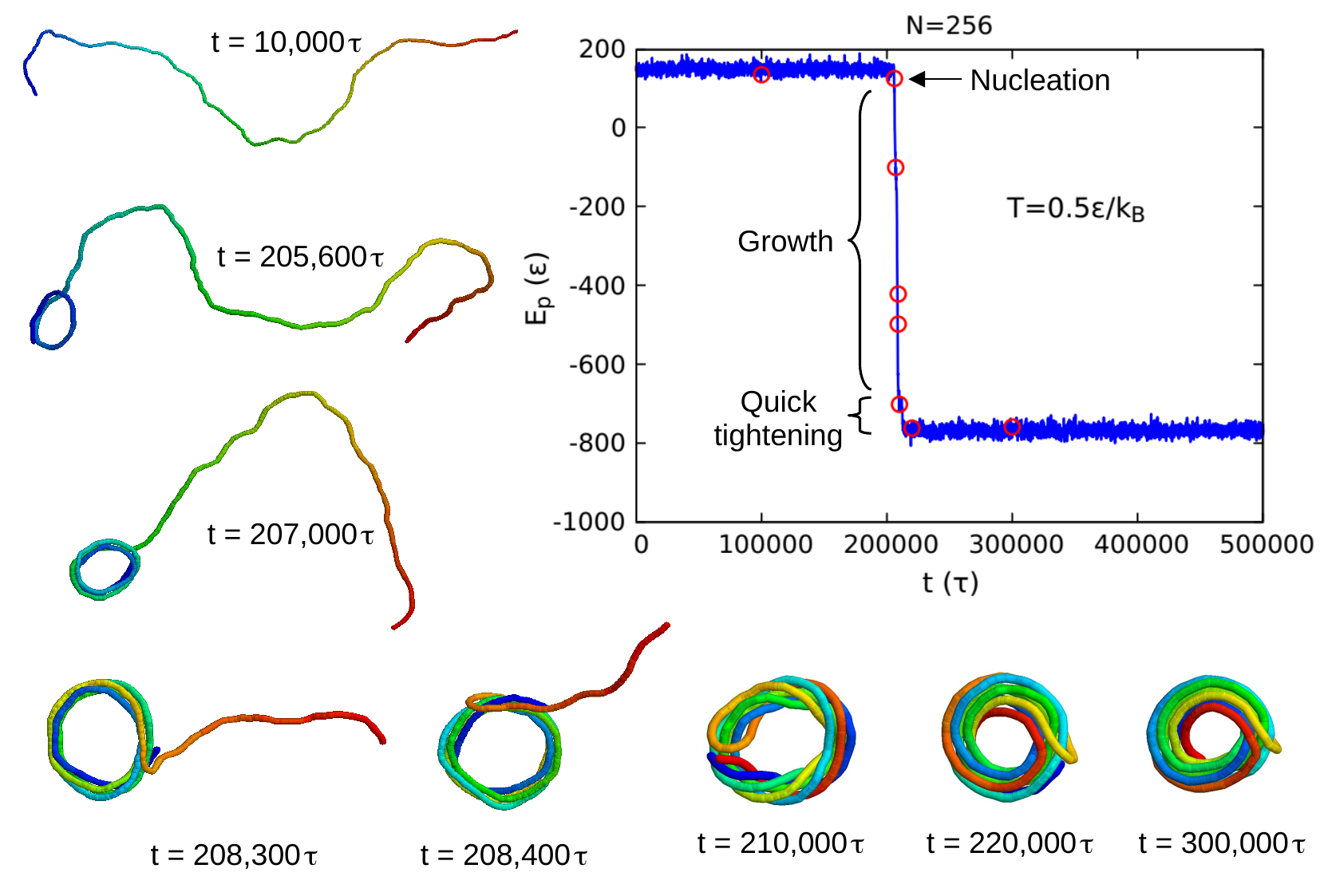}
\caption{A simulation trajectory showing the formation of a toroid with an
U-shaped region. The trajectory was obtained at temperature
$T=0.5\,\epsilon/k_B$ for the stiff polymer of $N=256$ beads with the bending
stiffness $\kappa_b=22\epsilon$. The time dependence of the potential energy
$E_p$ and several conformations drawn from the trajectory at selected points
(open circles) are shown. Note that the U-shaped region appears in the
conformation at $t=208,300\tau$ during the growth process and remains in the
toroidal structure.}
\label{fig:6a30traj}
\end{figure}

\begin{figure}
\includegraphics[width=14cm]{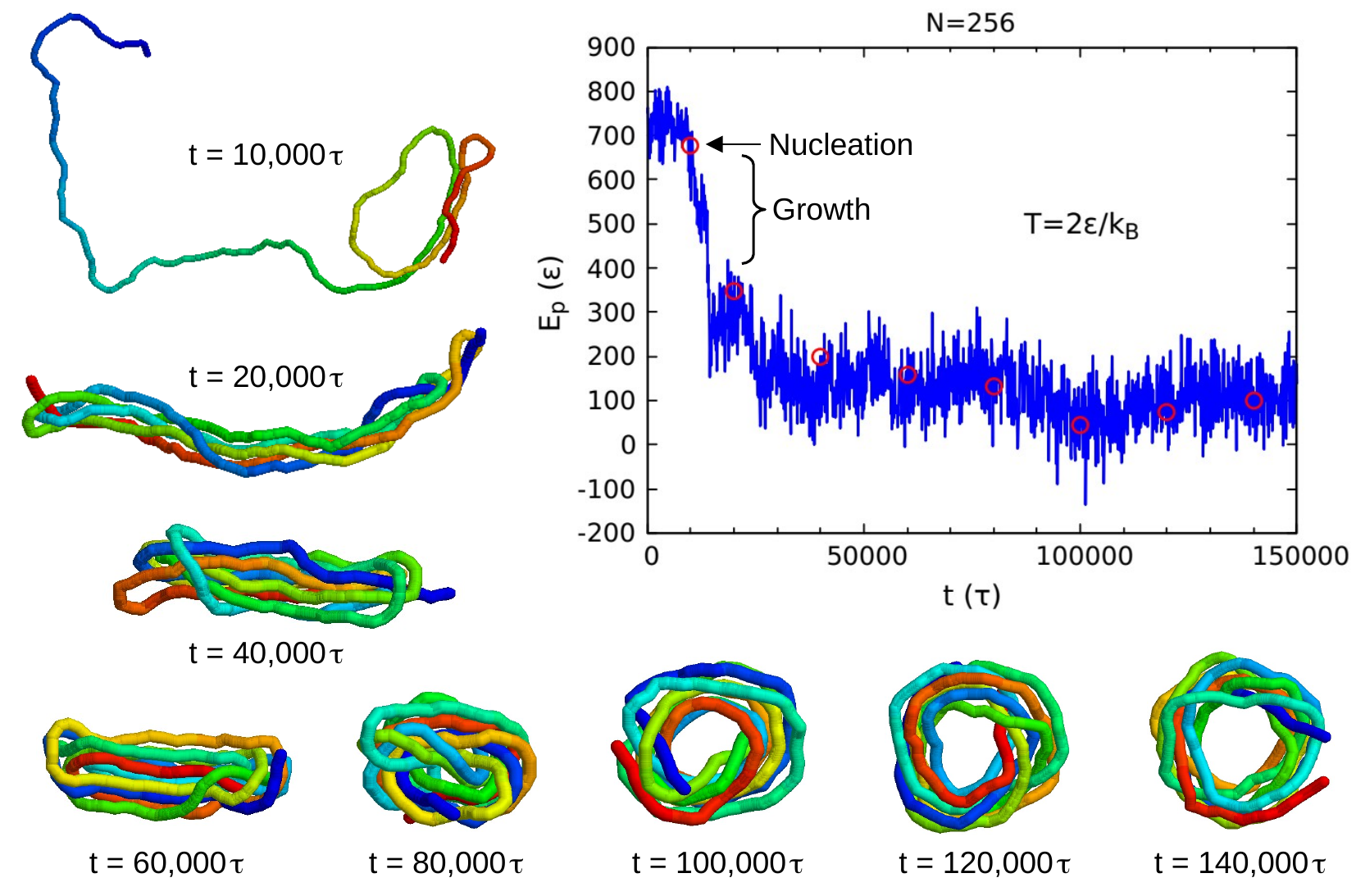}
\caption{
A simulation trajectory showing a conversion from a rod-like to a toroidal
conformation.  The trajectory was obtained at temperature
$T=2\,\epsilon/k_B$ for the stiff polymer of $N=256$ beads with the bending
stiffness $\kappa_b=22\epsilon$. The time dependence of the potential energy
$E_p$ and several conformations drawn from the trajectory at selected points
(open circles) are shown. The toroid structures in this trajectory are more
disordered than those in Fig.~S5 due to stronger thermal fluctuations. }
\label{fig:8a31traj}
\end{figure}

\end{document}